\documentclass{elsart}

\usepackage{natbib}
\usepackage{hyperref}

\usepackage{graphicx}

\usepackage{amssymb}
\usepackage{amsmath}
\usepackage{marvosym}

\def\st{$^{st}$}
\def\nd{$^{nd}$}
\def\rd{$^{rd}$}
\def\th{$^{th}$}
\def\deg{$^\circ$}

\begin{document}

\begin{frontmatter}

  \title{Interferometry Basics in Practice: Exercises}

  \author[MPIFR]{Florentin Millour},
  \author[Obs. Geneve]{Damien S\'egransan},
  \author[LAOG]{Jean-Philippe Berger},
  \author[LAOG]{Gilles Duvert} and
  \author[LAOG]{Fabien Malbet}

  \address[MPIFR]{Max-Planck-Institut f\"ur Radio-astronomie, Auf dem
    H\"ugel 69, 53121 Bonn, Germany}
  \address[Obs. Geneve]{Observatoire Astronomique de l'Universit\'e de
    Gen\`eve, 51, chemin des Maillettes, CH-1290 Sauverny, Switzerland}
  \address[LAOG]{Laboratoire d'Astrophysique, Observatoire de
    Grenoble, 414, Rue de la Piscine, Domaine Universitaire, 38400
    Saint-Martin d'H\`eres}

  \maketitle

  \begin{abstract}
    The following exercises aim to learn the link between the object
    intensity distribution and the corresponding visibility curves of
    a long-baseline ptical interferometer. They are also intended to
    show the additional constraints on observability that an
    interferometer has.

    This practical session is meant to be carried out with the \texttt{ASPRO}
    software, from the Jean-Marie Mariotti Center, but can also be
    done using other observation preparation software, such as viscalc
    from ESO.

    There are two main parts with series of exercises and the
    exercises corrections. The first one aims at understanding the
    visibility and its properties by practicing with simple examples,
    and the second one is about $UV$ coverage.
  \end{abstract}

  \begin{keyword}
    Optical long baseline interferometry, visibility, phase, $UV$
    coverage, VLTI, \texttt{ASPRO}

  \end{keyword}

\end{frontmatter}

\section{From model to visibility (exercises)}

This first series of exercises is made for training your
practical comprehension of the link between the image space (where you
usually work) and the Fourier space (where an interferometer produces
its measurements). Optical long-baseline interferometry experts
constantly switches between Fourier space and image
space. This training will help one get used to this
continuous switching.

\subsection*{What you will need for this particular practice session}

All exercises of this practice session will be done assuming a
simple yet unrealistic $UV$ coverage. Synthetic $UV$ tables
simulating such $UV$ coverages are provided for use:\\
\texttt{strip-[J,H,K,N]-[000,..,170].uvt}, where the number
\texttt{xxx} is the projected baseline angle in degrees. 

\texttt{ASPRO} is able to do many things, but this session will focus on the
\emph{MODEL/FIT}~-~\emph{UV plots and source modeling} part.

You will probably also need tables~\ref{tab4} and \ref{tab5}.

\begin{table}[htbp]
  \centering
  \caption{
    \footnotesize{
      All integrated \texttt{ASPRO} models one can use
      during this practice session and their useful parameters.
    }
  }
  \label{tab4}
  \bigskip
  \small
  \begin{tabular}{|l|l|l|}
    \hline 
    \texttt{ASPRO} name & Source shape & useful parameters\\
    \hline
    POINT      & Unresolved (Point source)              & None\\
    C\_GAUSS   & Circular Gaussian & FWHM Axis\\
    E\_GAUSS   & Elliptic Gaussian  & FWHM Axis (Major and Minor), Pos Ang\\
    C\_DISK    & Circular Disk             & Diameter\\
    E\_DISK    & Elliptical Disk& Axis (Major and Minor), Pos Ang\\
    RING       & Thick Ring                      & Inner Ring Diameter, Outer ring diam\\
    U\_RING    & Thin Ring            & Diameter (1 value: unresolved!)\\
    EXPO       & Exponential brightness    & FWHM Axis\\
    POWER-2    & B = 1/r$^{2}$                 & FWHM Axis\\
    POWER-3    & B = 1/r$^{3}$                & FWHM Axis\\
    LD\_DISK   & Limb-Darkened Disk        & Diameter, 'cu' and 'cv'\\
    BINARY     & Binary                    & Flux ratio, rho, theta\\
    \hline 
  \end{tabular}
\end{table} 

\begin{table}[htbp]
  \centering
  \caption{
    \footnotesize{
      Description of the parameters of the \texttt{ASPRO} model
      software module. {\bf Be sure to regard the parameter units.}
    }
  }
  \label{tab5}
  \bigskip
  \small
  \begin{tabular}{|l|l|l|l|l|l|l|l|}\hline 
    name / parameter & 1\st  & 2\nd  & 3\rd  & 4\th  & 5\th  & 6\th \\
    &  (") &  (") &  (no unit) &  (unit) &  (unit) &  (unit) \\
    \hline
    POINT     & R.A.& Dec& Flux &0&0&0\\
    C\_GAUSS  & R.A.& Dec& Flux & Diameter (") &0&0\\
    E\_GAUSS  & R.A.& Dec& Flux & Maj. diam. (") & Min. diam. (") & Pos. Ang. (\deg)\\
    C\_DISK   & R.A.& Dec& Flux & Diameter (") &0&0\\
    E\_DISK   & R.A.& Dec& Flux & Maj. diam. (") & Min. diam. (") & Pos. Ang. (\deg)\\
    RING      & R.A.& Dec& Flux & In diam. (") & Out diam. (") &0\\
    U\_RING   & R.A.& Dec& Flux & Diameter (") &0&0\\
    EXPO      & R.A.& Dec& Flux & Diameter (") &0&0\\
    POWER-2   & R.A.& Dec& Flux & Diameter (") &0&0\\
    POWER-3   & R.A.& Dec& Flux & Diameter (") &0&0\\
    LD\_DISK  & R.A.& Dec& Flux & Diameter (") & cu& cv\\
    BINARY    & R.A.& Dec& Flux & Flux Ratio& Rho (") & Theta (\deg)\\
    \hline 
    \multicolumn{7}{p{0.9\hsize}}{  
      \footnotesize{  
        Notes: - for the binary model, the {\bf Flux Ratio} is
        F$_{{\rm secondary}}$ / F$_{{\rm primary}}$,
        and {\bf Rho} \& {\bf Theta} are the angular separation (")
        and position angle (degrees) of the  binary.
      }
    }\\ 
    \multicolumn{7}{p{0.9\hsize}}{
      \footnotesize{
        - {\bf R.A.} and  {\bf Dec} are
        usually set to zero while {\bf  Flux} is set to one.
      }
    }\\
    \multicolumn{7}{p{0.9\hsize}}{
      \footnotesize{
        - ``0'' means you have to fill the
        parameter value a with zero. Be careful to put a 0 instead of
        leaving it blank, otherwise the software will crash !
      }
    }\\
  \end{tabular}
\end{table}

\subsection*{\underline{Exercise 1:} The diameter of a star.}

The aim of this exercise is to have  first contact with uniform
disks, which are very often used to perform photospheric diameters
fits or first-order interpretations of the data.

\paragraph*{Plotting a uniform disk visibility curve:}
Given a star with a 2 milli-arc-second (mas) photospheric radius, use
the model function to plot the visibility versus the projected
baseline length (baseline \texttt{radius}). For this purpose, you can
use the \textit{MODEL/FIT.UV Plots \& Source Modeling} menu.

\paragraph*{Zero visibility:}
At what baseline does the visibility become equal to zero \citep[you can
refer to][for example]{2007NewAR..51..576B}? Use this
number to evaluate the disk diameter.

\paragraph*{Diameter uniqueness:}
Can you measure a unique diameter if your visibility is non-zero but
you know the star looks like a uniform disk? If yes, how?

\subsection*{\underline{Exercise 2:} Binary star.}

When the amount of data you get is a low number of visibilities, the
object's complexity for your interpretation cannot be too
high. Therefore, binary models are often used to understand the data
when asymmetries happen to be proved by means of interferometry (by a
non-zero closure phase) or by indirect clues (a polarization  of the
target, for example). Therefore, one has to understand the behavior of
such a model.

\paragraph*{Plotting a binary star visibility curve:}
Display the visibility and phase as a function of projected baseline
(using the file \texttt{strip-K-60}) of a binary with unresolved
components with 4\,mas separation, a flux ratio of 1, and a position angle
of 30 degrees. Do the same thing with different separations. Comment
on the result.

\paragraph*{Varying the flux ratio:}
Using the previous model, now vary the flux ratio from 1 to 1e-6.
Comment on how the dynamic range requirement to detect the companion
translates into visibility and phase constraints?

\paragraph*{Phase versus visibility:}
Would the phase alone be sufficient to constrain the binary
parameters?

\subsection*{\underline{Exercise 3:} Circumstellar disk.}

The last model we will see in this practice session is a Gaussian disk
that can, in a first approximation, simulate a circumstellar disk, or an
optically thick stellar wind. The idea here is to understand how
visibilities change with source elongation.

\paragraph*{Plotting a Gaussian disk visibility curve:}
Display the visibility curve of a disk which is assumed to have an
elliptical Gaussian shape (model \texttt{E\_GAUSS}). Use the minor and
major axes (parameters 4 \& 5) to simulate an inclination. The display
should be done for several PAs (\texttt{strip-*-(0,30,45,60,90)}).

\paragraph*{Aspherity and visibility variations:}
Comment on how the aspherity induced by the inclination changes
the visibility function at a given projected angle.

\subsection*{\underline{Exercise 4:} Model confusion and accuracy.}

If the baseline you have chosen is too long or too small relative to
the typical size of your source, this may cause problems when you try
to interpret your data. Here you will see why.

\paragraph*{Plotting several model visibilities:}
Use the model function to compute the visibility of a star with a
uniform disk brightness distribution (2\,mas radius), circular
Gaussian disk (1.2 mas radius), and binary (flux ratio 1, 1 mas separation, 
$45^{\circ}$ PA) with the baseline stripe \texttt{strip-K-60}. No
superposition of the plot is possible, so use the \emph{show plot in
  browser} option, save it as a postscript file, and compare the
different files afterwards. Compare their visibilities at 100\,m in the
K band.

\paragraph*{Model confusion at small baselines:}
How can we distinguish between these various models?
What about measurements at 200 m? What do you conclude?

\paragraph*{The role of measurement accuracy:}
Does the measurement accuracy play a role in such model discrimination?

\paragraph*{Which baseline for which purpose:}
Construct a 2-component model in which a central, unresolved star
(\texttt{POINT}) is surrounded by an inclined, extended structure. You
can use an elliptical Gaussian distribution (\texttt{E\_GAUSS}) for
this purpose (minor and major axes in the range 0.5 to 15 mas).

Try two scenarios:
\begin{itemize}
\item an extended source easily resolvable but with a flux
  contribution much smaller than the star;
\item a smaller extended source but with a larger flux contribution.
\end{itemize}
What are the best baseline lengths for estimating the size and relative
flux contributions with an interferometer?

\subsection*{\underline{Exercise 5:} Choosing the right baselines.}

Given a specific object's shape, one can determine how a baseline
constrains a given model parameter. We will see this aspect here.
In order to determine the parts of the $UV$ plane which constrain
the model most, one can make use of the first derivative of the
visibility with respect to a given parameter (e.g. derivative of
visibility versus diameter).

\paragraph*{Uniform disk:}
Choose a uniform disk model. What is the most constraining part in the
$UV$ plane?

For this exercise, in the \texttt{UV EXPLORE} panel, use \texttt{V} versus
\texttt{U}, check the \texttt{under-plot model image} option, and
choose the appropriate derivative in the \texttt{plot what...} line.

\paragraph*{Gaussian disk:}
Do the same exercise using a Gaussian disk.

\subsection*{\underline{Exercise 6:} An unknown astrophysical object.}

The wavelength at which an object is observed is also important. This
exercise attempts to illustrate this point.

\paragraph*{Loading and displaying a home-made model:}
Load the fits table \texttt{fudisk-N.fits} corresponding to the
simulation of a certain type of astrophysical object (here, a disk
around an FU Orionis object) using \textit{OTHER/Display a GDF or FITS
  image} menu. If the color scale is not appropriate, check the
\textit{Optional parameters} button and select another color
scale. Notice what the contrast of the object (angular
units in radians) is.

\paragraph*{Computing the visibilities of a home-made model:}
Compute the visibility of the model in the N-Band with 
\textit{MODEL/ FIT.UV Plots \& Source Modeling/USE HOMEMADE MODEL}. To
do so, select the appropriate grid \texttt{strip-N-60} in the
\textit{Input Information} menu. Use  \texttt{UV EXPLORE} to plot the
visibility amplitude versus the spatial frequency radius.

\paragraph*{Comparing visibilities for different wavelengths:}
Repeat these operations in the K band, then the H and J ones.
Compare the visibility profiles. Conclude on the optimal wavelength to
observe the object with the VLTI (maximum baseline is 130\,m today).

\subsection*{\underline{Exercise 7:} Play with spectral variations,
  closure phases, etc.}

Bonus exercise: Try to guess what this model shows just by looking at
the visibilities (please do not cheat!).

\paragraph*{Plotting visibilities and closure phase versus wavelength:}
Using a given model of a binary star ($\gamma^2$ Vel, file
gammaVelModelForAspro.fits), try to plot visibility and closure phases
as a function of wavelength (see the \emph{OTHER ... Export UV table as OI
  fits} and \emph{OTHER ... OI fits file explorer} menus).

\paragraph*{Qualitative understanding:}
Compare the obtained visibilities with what you would get with an
\texttt{ASPRO} model of one Gaussian and one uniform disk of diameters 0.5 mas,
a separation of 3.65 mas, and an angle of 75 degrees. What do you
conclude?

\paragraph*{Looking at the solution:}
After your conclusions, you can  look at the model by opening it
with a fits viewer (for example,
fv\footnote{\url{http://heasarc.gsfc.nasa.gov/lheasoft/ftools/fv/}}).

\newpage

\section{From model to visibility (Correction)}

The author of this paper carried out the previous exercises using
\texttt{ASPRO} (and an image processing software for superposition of the
graphs: \texttt{GIMP}) to give an idea of what one should get with \texttt{ASPRO} when
following the previous exercises. {\bf Please try to do the exercises
  yourself before reading these corrections}.

\subsection*{\underline{Exercise 1:} The diameter of a star.}

\begin{figure}[htbp]
  \centering
  \begin{tabular}{cc}
    \includegraphics[width=0.47\textwidth]{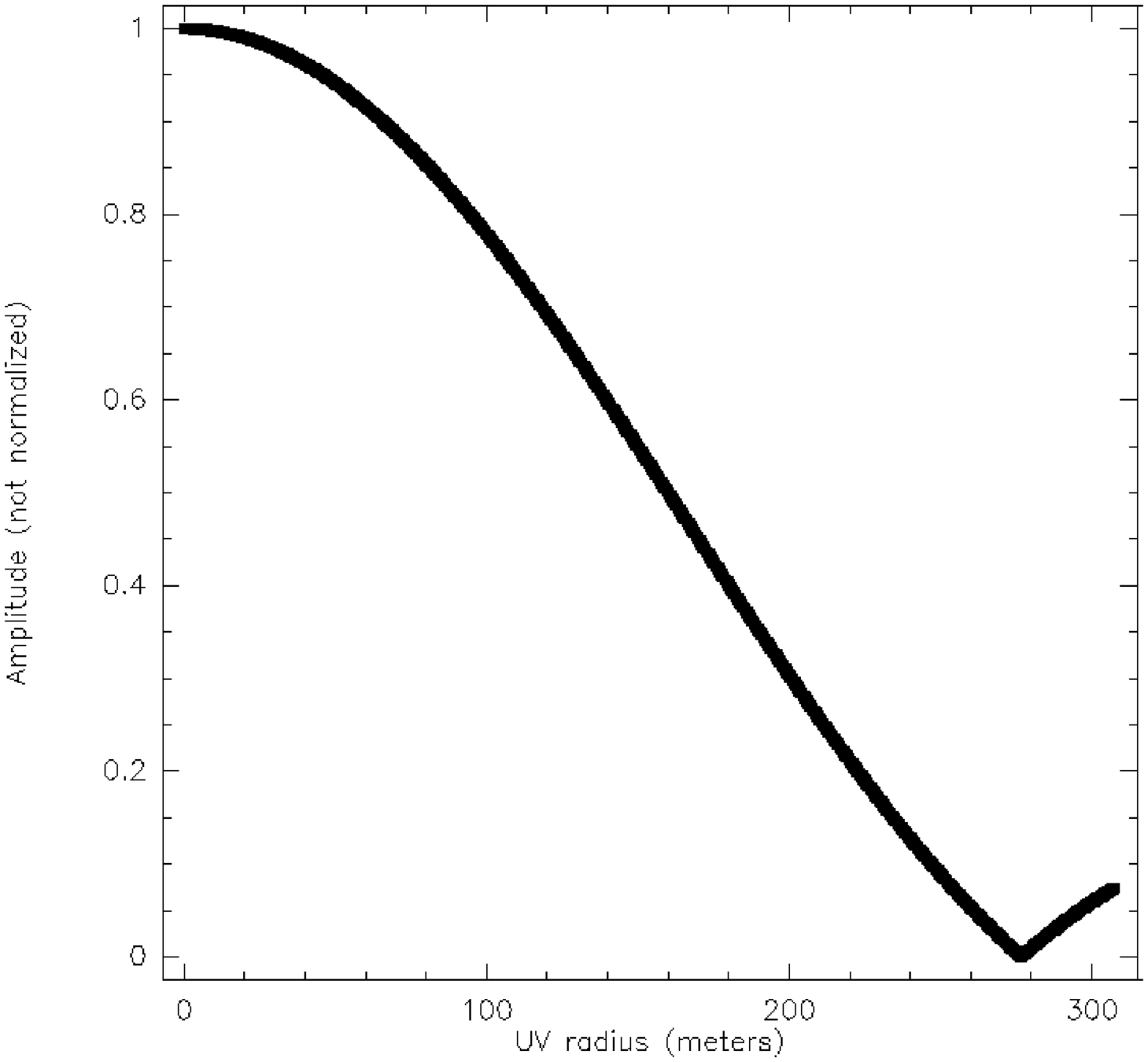}&
    \includegraphics[width=0.47\textwidth]{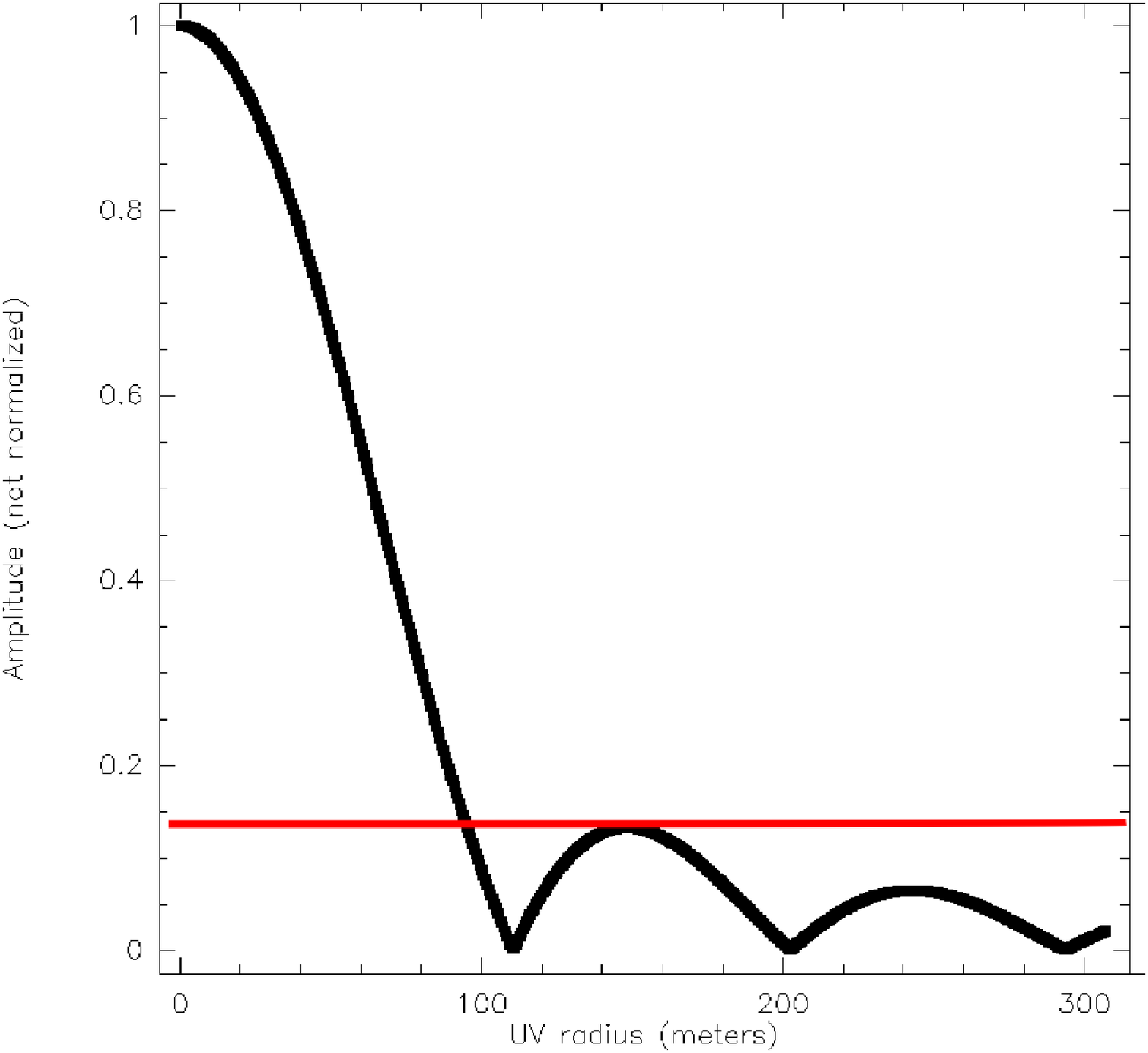}
  \end{tabular}
  \caption[Uniform disk visibility function cut versus base length.]{
    \footnotesize{
      \emph{\bf Left:} The uniform disk visibility function cut from
      exercise 1.
      \emph{\bf Right:} The place where one can infer a unique
      diameter from a single visibility measurement for a uniform
      disk is where the measured contrast is above $0.15$. Indeed,
      below $0.15$, several diameters can be inferred from this single
      visibility measurement (as shown by the red line).
    }
  }
  \label{fig:unfiDiskVisibility}
\end{figure}

\paragraph*{Plotting a uniform disk visibility curve:}
The figure produced by \texttt{ASPRO} should look like
Fig.~\ref{fig:unfiDiskVisibility}, left.

\paragraph*{Zero visibility:}
The visibility for a uniform disk of diameter $a$ is given by the
following expression:

\begin{equation}
  V(\rho) = 2 \frac{\mbox{J}_1(\pi a \rho)}{\pi a
    \rho}
\end{equation}

$\rho = B / \lambda$ ($B$ and $\lambda$ in meters) being the spatial
frequency, $a$ being the star diameter (in radians), and $\mbox{J}_1$
the 1\st\ order Bessel function. Therefore, the value for which the
visibility becomes zero is $B = 1.22 \lambda / a$.

Here, the visibility zeroes around the 280\,m baseline. This gives
an approximately $1.89$\,mas diameter for the star, close to the 2\,mas
input.

\paragraph*{Diameter uniqueness:}
The right plot in Fig.~\ref{fig:unfiDiskVisibility} gives a hint of
the answer: There are parts of the visibility function which are
monotonic (above the red line). In these parts, one visibility gives
a unique solution to the diameter of the star. In the parts below the
red line, a given visibility corresponds to many different
solutions (as the line crosses several points of the
curve). Therefore, there is a lower limit on the visibility value ($V
\gtrsim 0.15$) where one can infer a unique diameter from a unique
measurement.


\subsection*{\underline{Exercise 2:} Binary.}

\begin{figure}[htbp]
  \centering
  \begin{tabular}{cc}
    \includegraphics[width=0.47\textwidth, angle=0]{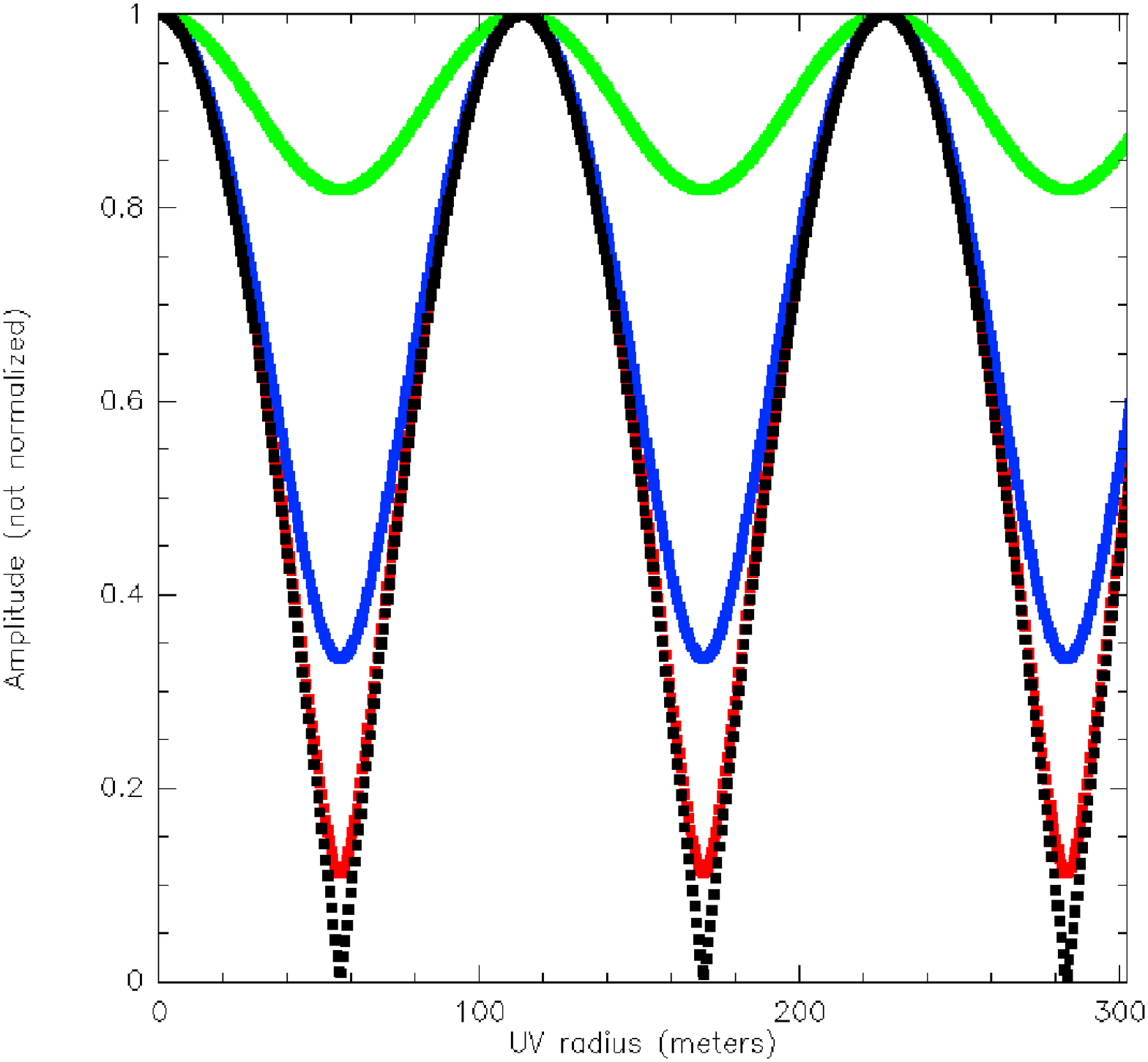}&
    \includegraphics[width=0.47\textwidth, angle=0]{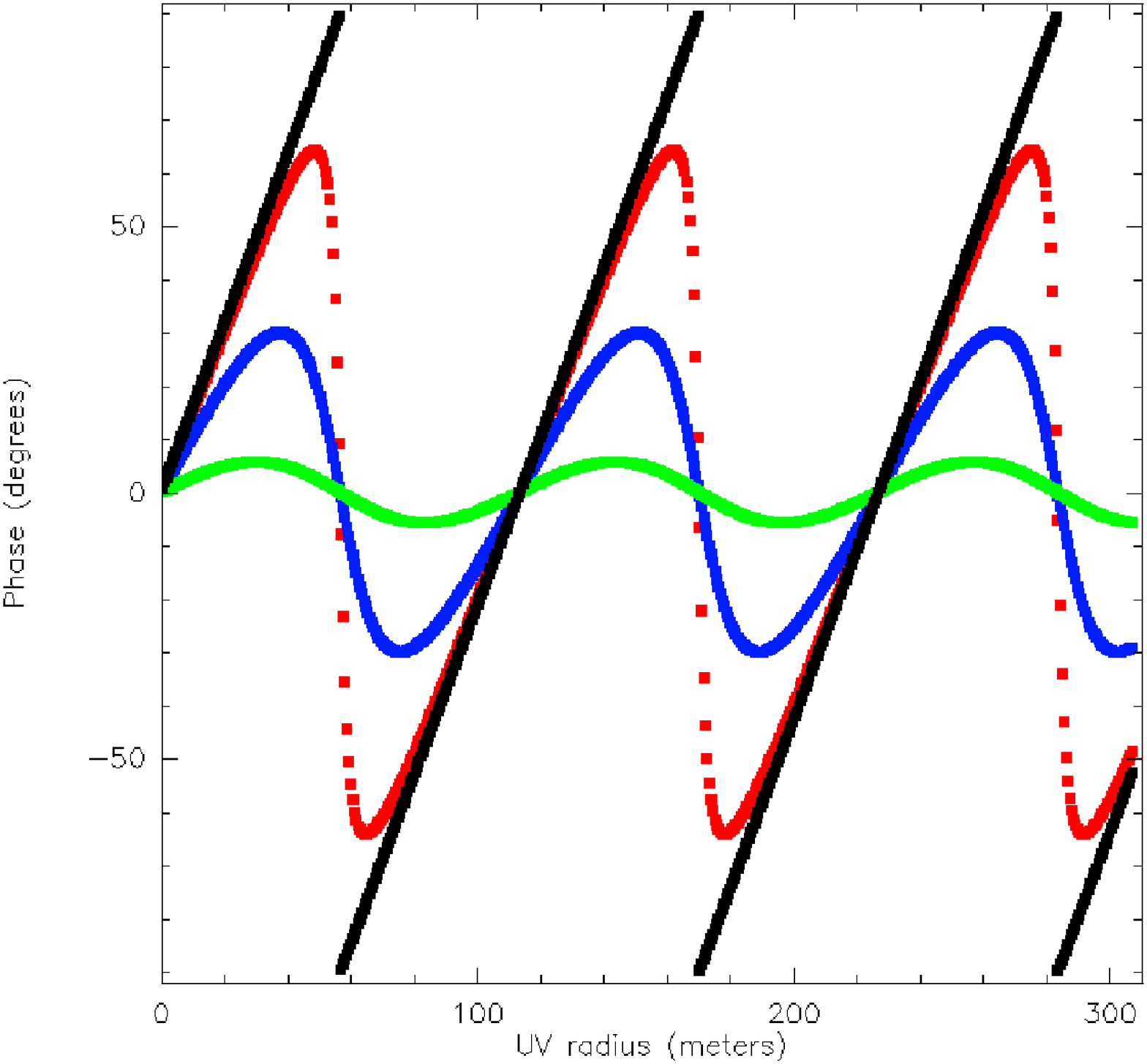}
  \end{tabular}
  \caption[Binary star visibility function cut versus base length.]{
    \footnotesize{
      \emph{\bf Left:} The visibility amplitude for a separation of
      4\,mas and different flux ratios (from the lower to the upper
      curves: 1 to 1 in black, 0.8 to 1 in
      red, 0.5 to 1 in blue, and 0.1 to 1 in green).
      \emph{\bf Right:} The visibility phase for the same separation
      and flux ratios.
    }
  }
  \label{fig:visiBinary1}
\end{figure}

\paragraph*{Plotting a binary star visibility curve:}
The visibility and phase functions of a binary star are periodic since
the image is made of Dirac functions. One can see what can be expected
for a  4\,mas-separation binary star in
Fig.~\ref{fig:visiBinary1}. One can see that the visibility does not
have a cosine shape, but has sharp changes at visibility 0 for a 1 to
1 binary (black line).

\paragraph*{Varying the flux ratio :}
For other flux ratios (0.8 in red, 0.5 in blue, and 0.1 in green),
both the phase and visibility get smoother, and the contrast of the
variations gets dimmer. Please note that these are visibility plots
made with ``AMP'' and not ``AMP\^2'' in \texttt{ASPRO}.

\begin{figure}[htbp]
  \centering
  \begin{tabular}{cc}
    \multicolumn{2}{c}{\includegraphics[width=0.4\textwidth]{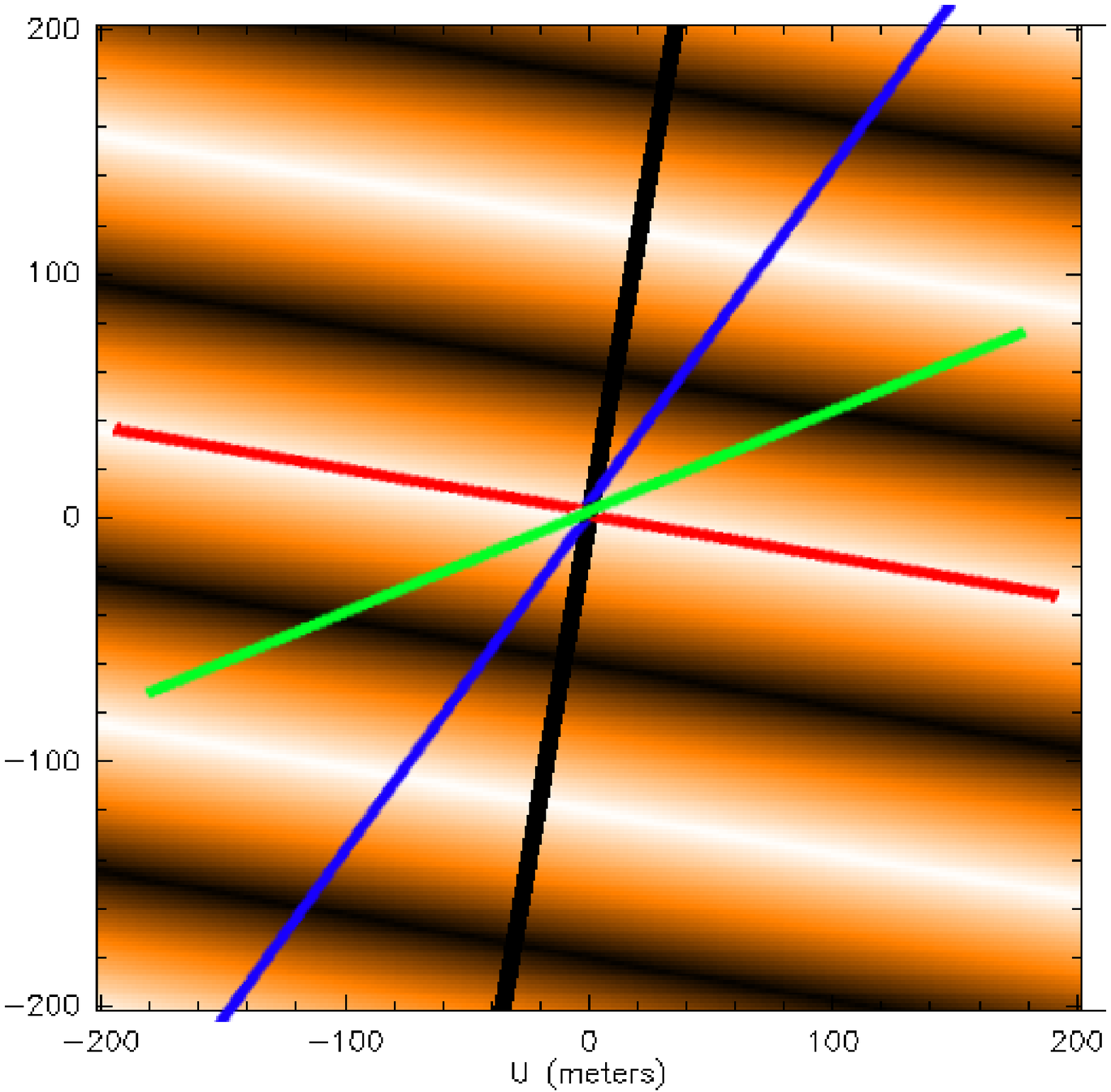}}\\
    \includegraphics[width=0.47\textwidth]{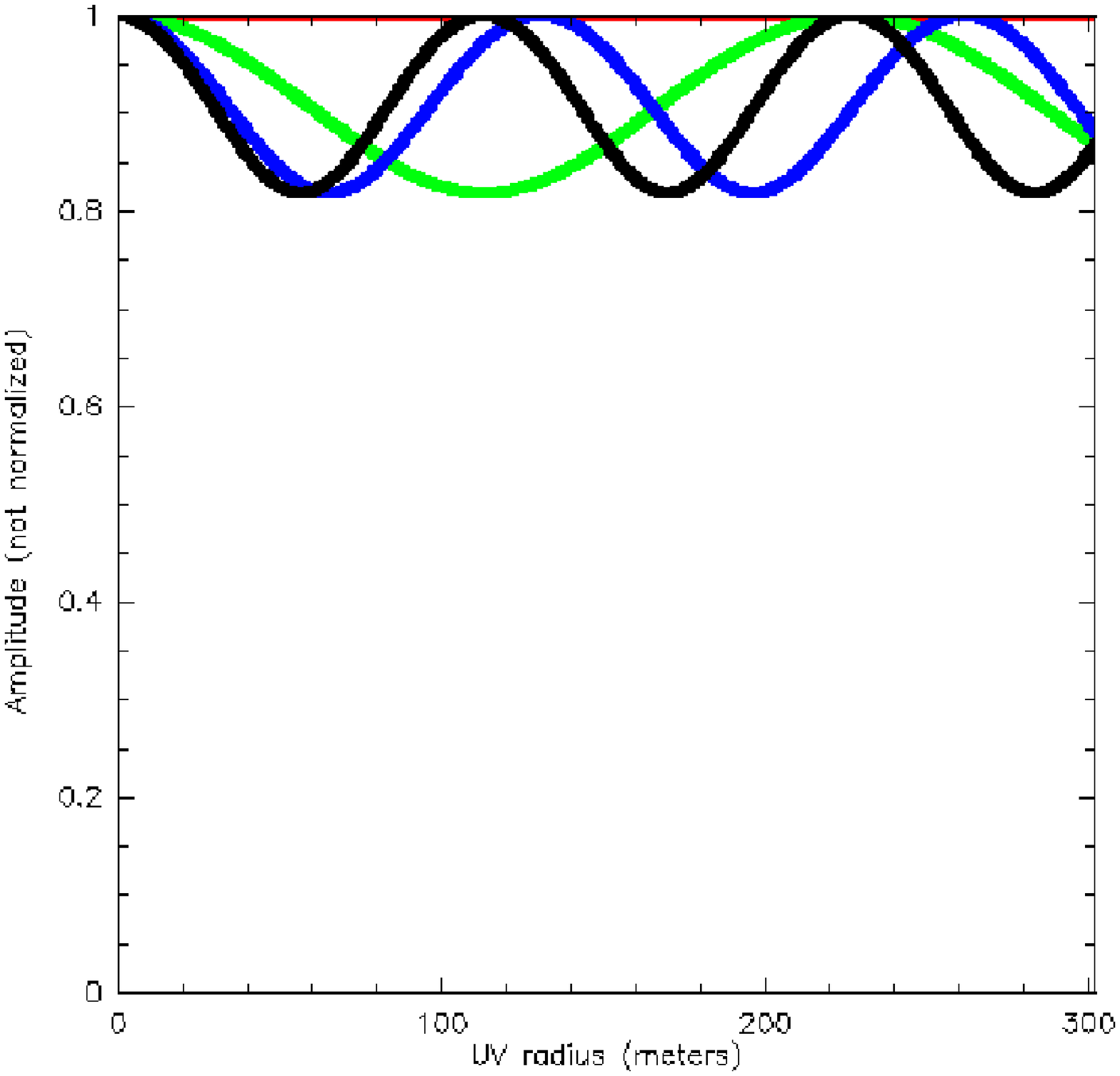}&
    \includegraphics[width=0.47\textwidth]{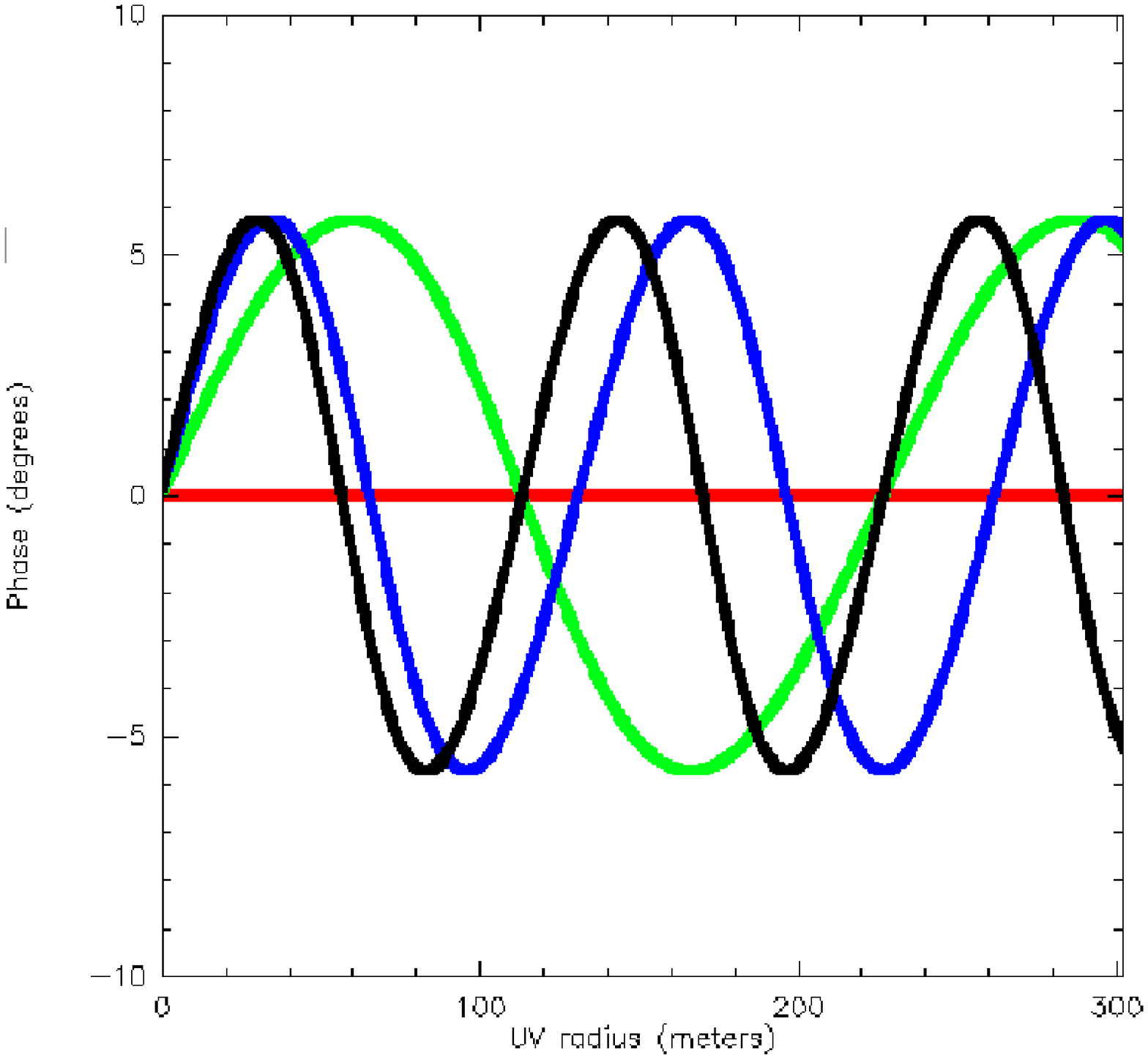}
  \end{tabular}
  \caption[Binary star visibility function cut versus position angle.]{
    \footnotesize{
      \emph{\bf Top:} $UV$ map showing the previous binary star
      visibility modulus. This visibility shows characteristic
      modulation stripes perpendicular to the binary orientation.
      \emph{\bf Bottom-Left:} Visibility modulus as a function of base
      length for the different projected stripes, in different
      colors (the number of sine arches increases with the different
      orientations): red is 90\deg\ relative to the binary orientation,
      green is 60\deg, blue is 30\deg\ and black is 0\deg.
      \emph{\bf Bottom-right:} Visibility phase as a function of base length
      for the same stripes as before.
    }
  }
  \label{fig:visiBinary2}
\end{figure}

\paragraph*{Phase versus visibility:}
To see how visibility or phase can constrain a binary star model, one
can just try to change the baseline orientation and see how visibility
and phase vary. In Fig.~\ref{fig:visiBinary2}, bottom-right, one can
see the result of such exercise. One has seen that the phase is
sensitive to the contrast of the binary (Fig.~\ref{fig:visiBinary1}),
as is the visibility, but it is also sensitive to the position angle
(both of the binary star and of the baseline) and the binary
separation. Therefore, the phase can be used instead of the visibility
to constrain the binary parameters!

\subsection*{\underline{Exercise 3:} Circumstellar disk.}

\begin{figure}[htbp]
  \centering
  \begin{tabular}{cc}
    \includegraphics[width=0.47\textwidth, angle=0]{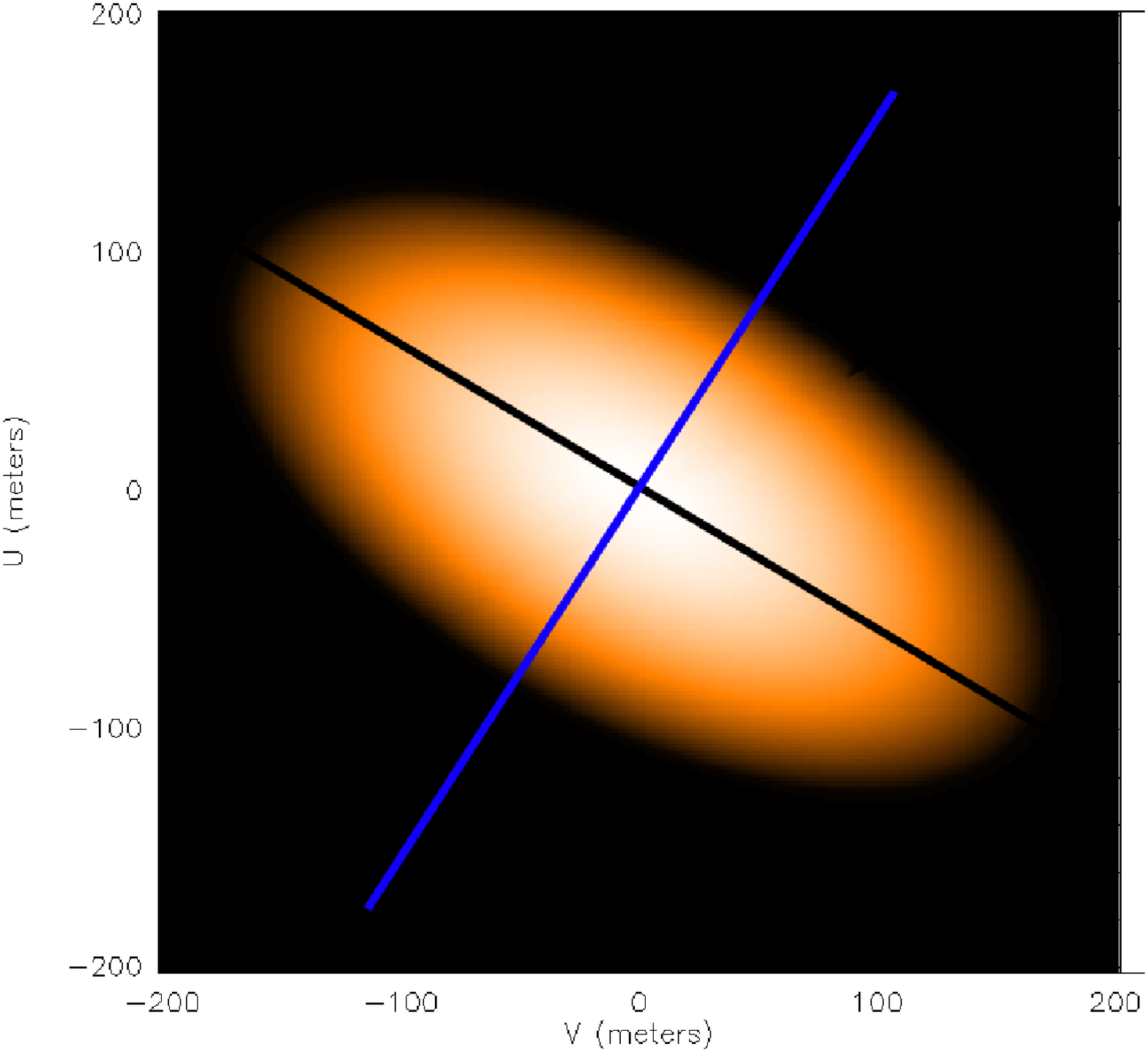}&
    \includegraphics[width=0.47\textwidth, angle=0]{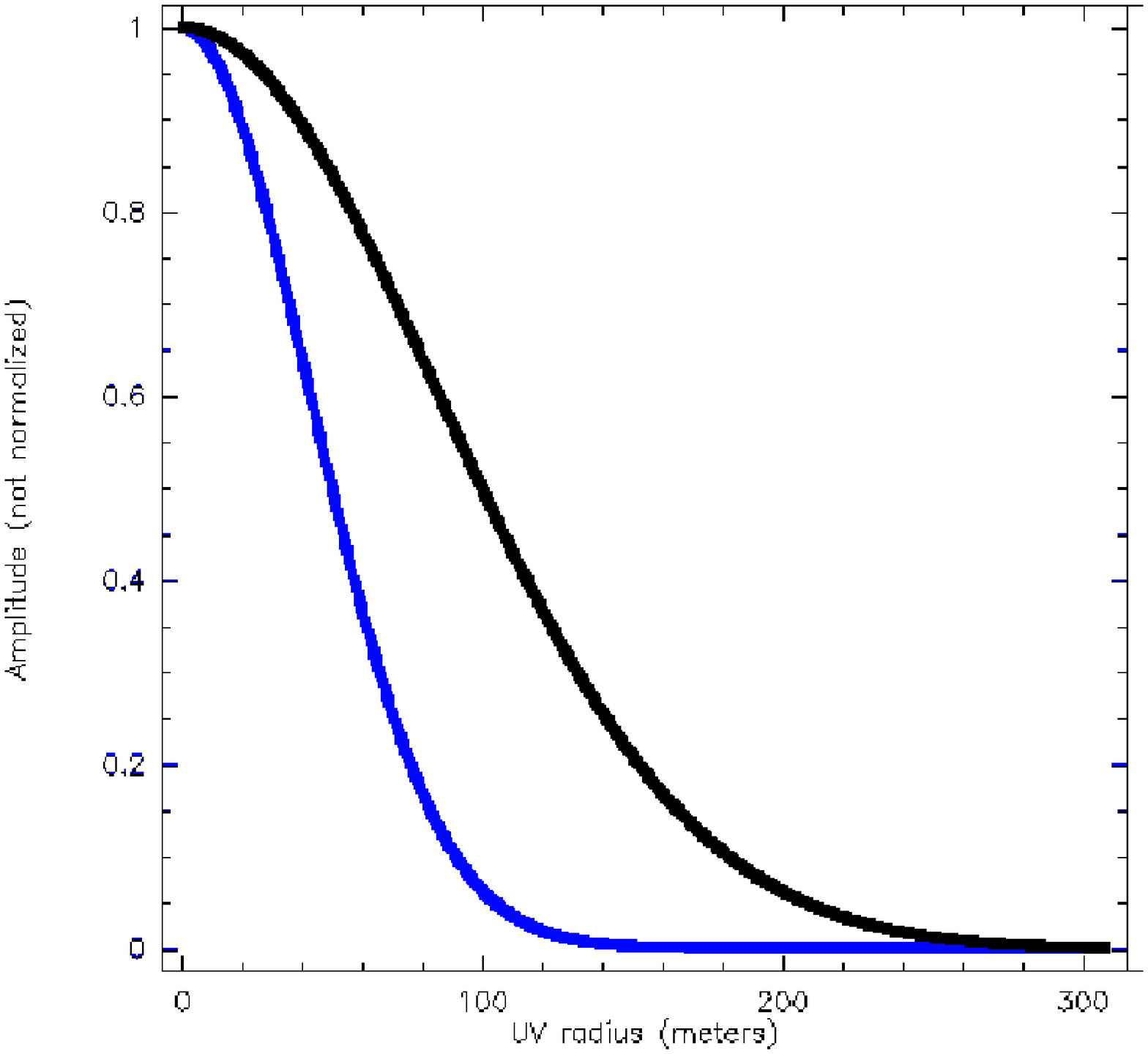}
  \end{tabular}
  \caption[Gaussian disk visibility function.]{
    \footnotesize{
      \emph{\bf Left:} $UV$ map showing the visibility amplitude for a
      2x4\,mas elongated Gaussian disk. The black and blue lines
      correspond to the right part of the figure.
      \emph{\bf Right:} Visibility versus base length for the minor
      (black, upper curve) and major (blue, lower curve) axes of the
      Gaussian disk. Please note
      that the visibility is lower in the direction of the major axis
      and greater in the minor axis direction.
    }
  }
  \label{fig:diskModel}
\end{figure}

\paragraph*{Plotting a Gaussian disk visibility curve:}
The example of this exercise is a disk of 2x4\,mas
and a position angle of 60 degrees. The resulting $UV$ map and cuts
in the $UV$ plane are shown in Fig.~\ref{fig:diskModel}.  

\paragraph*{Aspherity and visibility variations:}
As one can see, for a given baseline and a varying position angle, the
visibility goes up and down, the minimum corresponding to the major
axis and the maximum corresponding to the minor axis. One has to note
(and can check) that the phase function for such a model is zero.

\subsection*{\underline{Exercise 4:} Model confusion and accuracy.}

\begin{figure}[htbp]
  \centering
  \includegraphics[width=0.47\textwidth, angle=0]{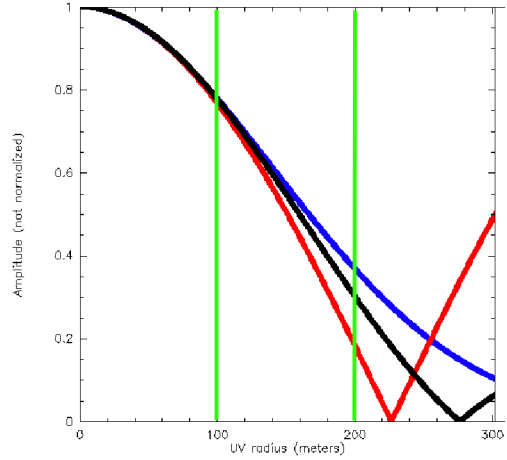}
  \caption[Model confusion, and baseline range.]{
    \footnotesize{
      Illustration of model confusion for different baselines and
      models: blue (top curve) is a Gaussian disk, red (bottom curve)
      is a binary star, and black (middle curve)
      is a uniform disk. The first green vertical line represents a
      100\,m baseline. If the accuracy is not better than about 0.01,
      then no distinction can be made between the different
      models. For the 200\,m baseline (second green vertical line), an
      accuracy of 0.1 or better is sufficient to discriminate between
      the different models. Note, however, that one needs more than 1
      point to be able to really distinguish between the models.
    }
  }
  \label{fig:modelConfusion}
\end{figure}

\paragraph*{Plotting several model visibilities:}
In Fig.~\ref{fig:modelConfusion} different plots are superimposed
using the \texttt{GIMP} software. The uniform disk is in black, the Gaussian in
blue, and the binary in red.

\paragraph*{Model confusion at small baselines:}
Two green lines represent the 100\,m and 200\,m baselines. One can see
the importance of multi-measurements at 
different baselines to be able to disentangle the different objects'
shapes. With only one visibility measurement, one will never be able
to distinguish between these different models. With the two visibility
measurements, one can see that it will be possible to disentangle the
different models if both very different baselines AND a sufficient
accuracy can be reached.

\begin{figure}[htbp]
  \centering
  \begin{tabular}{cc}
    \includegraphics[width=0.47\textwidth,
    angle=0]{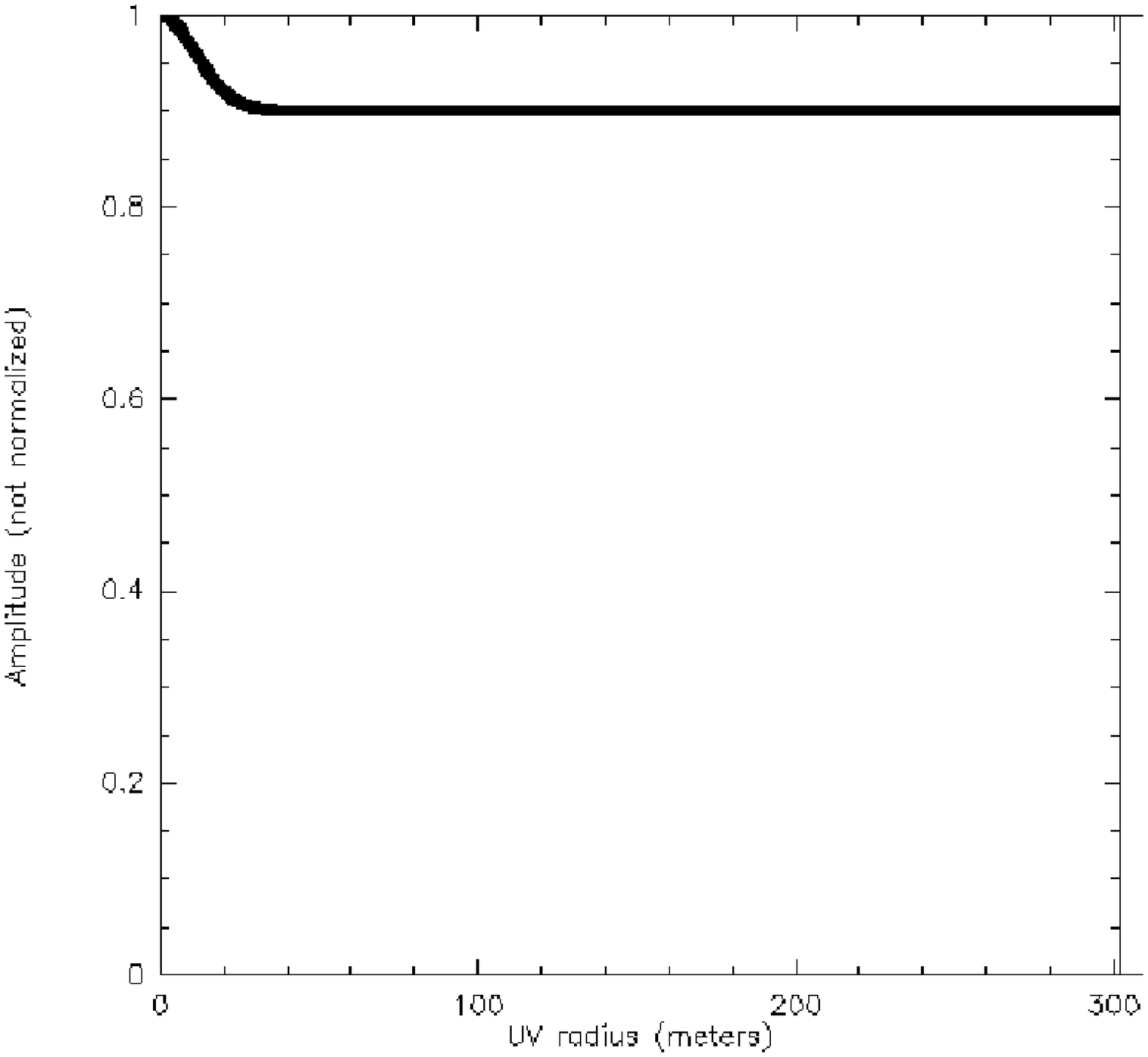}&
    \includegraphics[width=0.47\textwidth,
    angle=0]{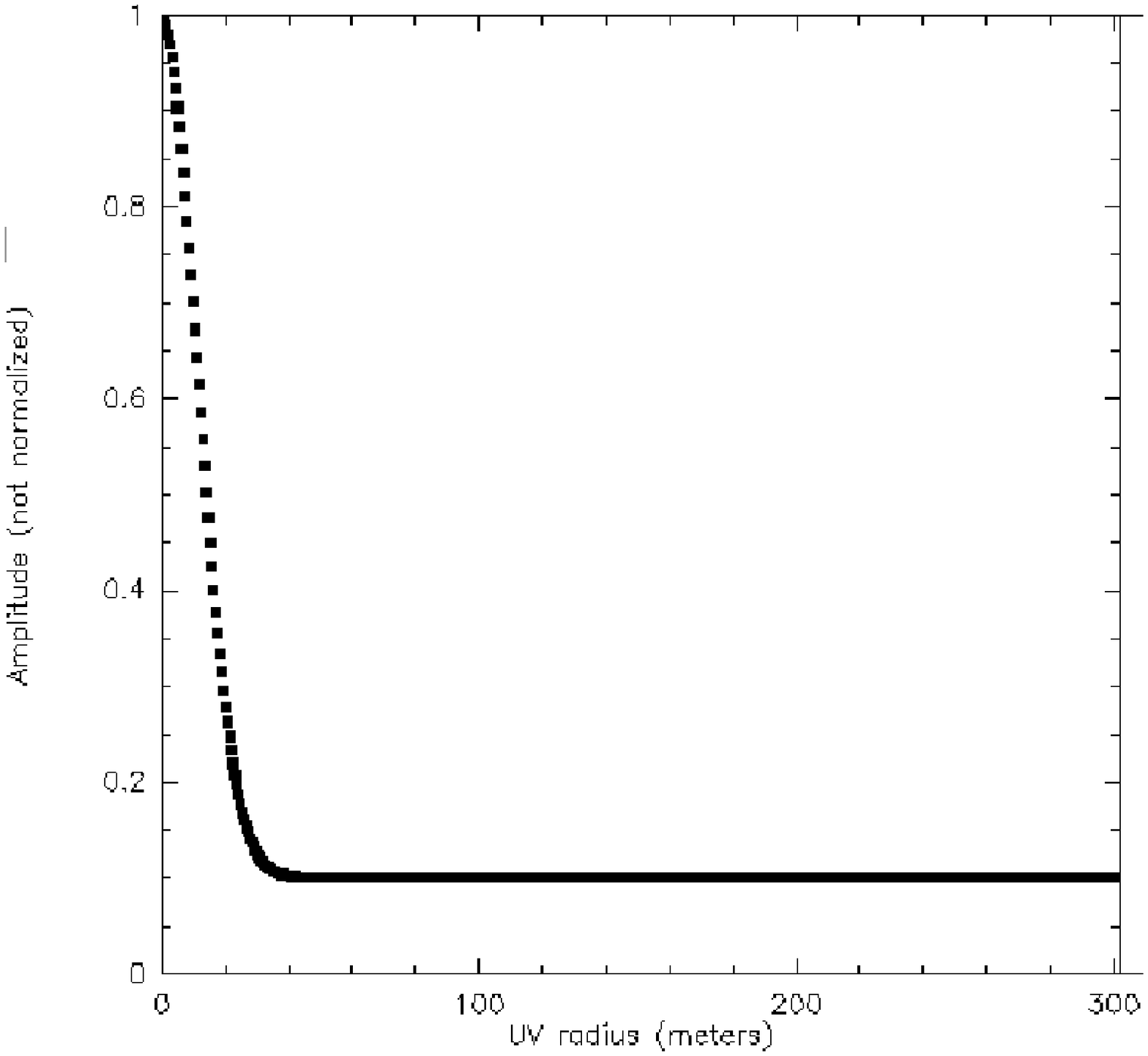}
  \end{tabular}
  \caption[Model confusion, and accuracy.]{
    \footnotesize{
      \emph{\bf Left:} Visibility cut for a point star accounting for
      90\% of the total flux and a 15\,mas Gaussian disk accounting for 10\%
      of the flux.
      \emph{\bf Right:} Visibility cut for a point star accounting for
      10\% of the total flux and a 15\,mas Gaussian disk accounting for 90\%
      of the flux.
    }
  }
  \label{fig:modelConfusion2}
\end{figure}

\paragraph*{The role of measurement accuracy:}
Here, if one has visibility error bars of 0.05, one can distinguish
between the sources using the 100 and 200\,m  baselines, but not with
an accuracy of 0.3. This importance of accuracy in model confusion is
illustrated by Fig.~\ref{fig:modelConfusion2}. The 1st plot (on the
left) is a central point with 90\% of the total flux contribution and
a large (15mas) Gaussian disk around, accounting for 10\% of the
flux. Errors bars of 0.1 would prevent one from finding the Gaussian
component, and the measured visibility ($0.9\pm0.1$) would be
compatible with a completely unresolved star. The right graph shows
the same but with the flux ratio inverted (90\% of the flux to the
disk and 10\% to the central star). In this case, the visibilities
would be compatible with a fully resolved, extended component, given an
accuracy of 0.1 on the visibility, and the point source (the central
star) would not be detected. So, not only the number of baselines but
also the accuracy of the measurement is of high importance to
distinguish between different models.

\paragraph*{Which baseline for which purpose:}
Here, one can see that since there is an unresolved source in the object
image (a central source), measuring the visibility at long baseline
directly provides the unresolved$_{\rm flux}$/resolved$_{\rm
  flux}$  flux ratio.

If one wants to get information on the disk itself (size, shape, etc.),
the shorter baselines are more appropriate, since the visibility will vary
according to the source shape.

\subsection*{\underline{Exercise 5:} Choosing the right baselines.}

The idea here is to use the nice feature of \texttt{ASPRO} that is able to plot
the derivatives of visibility versus the different parameters of the
input model. To do so, one has to go to the \emph{UV explore} panel,
select U as X data and V as Y data, and finally select $d(AMP)/d(4)$
in the \emph{Plot what...} part. This will plot  the derivative of the
visibility amplitude versus the 4\th\ parameter (radius, as shown in table
\ref{tab5}).

\begin{figure}[htbp]
  \centering
  \begin{tabular}{cc}
    \includegraphics[width=0.47\textwidth, angle=0]{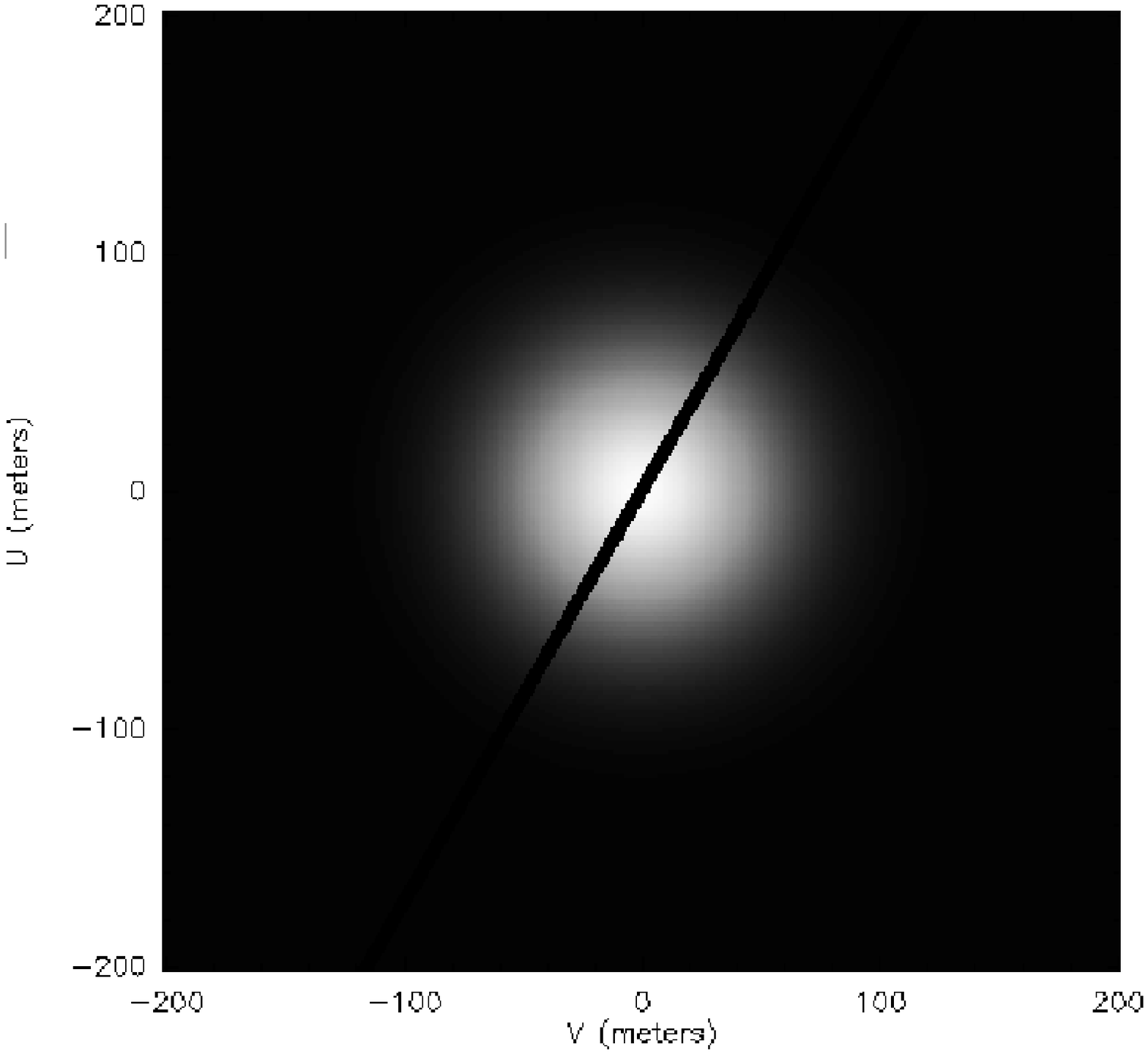}&
    \includegraphics[width=0.47\textwidth, angle=0]{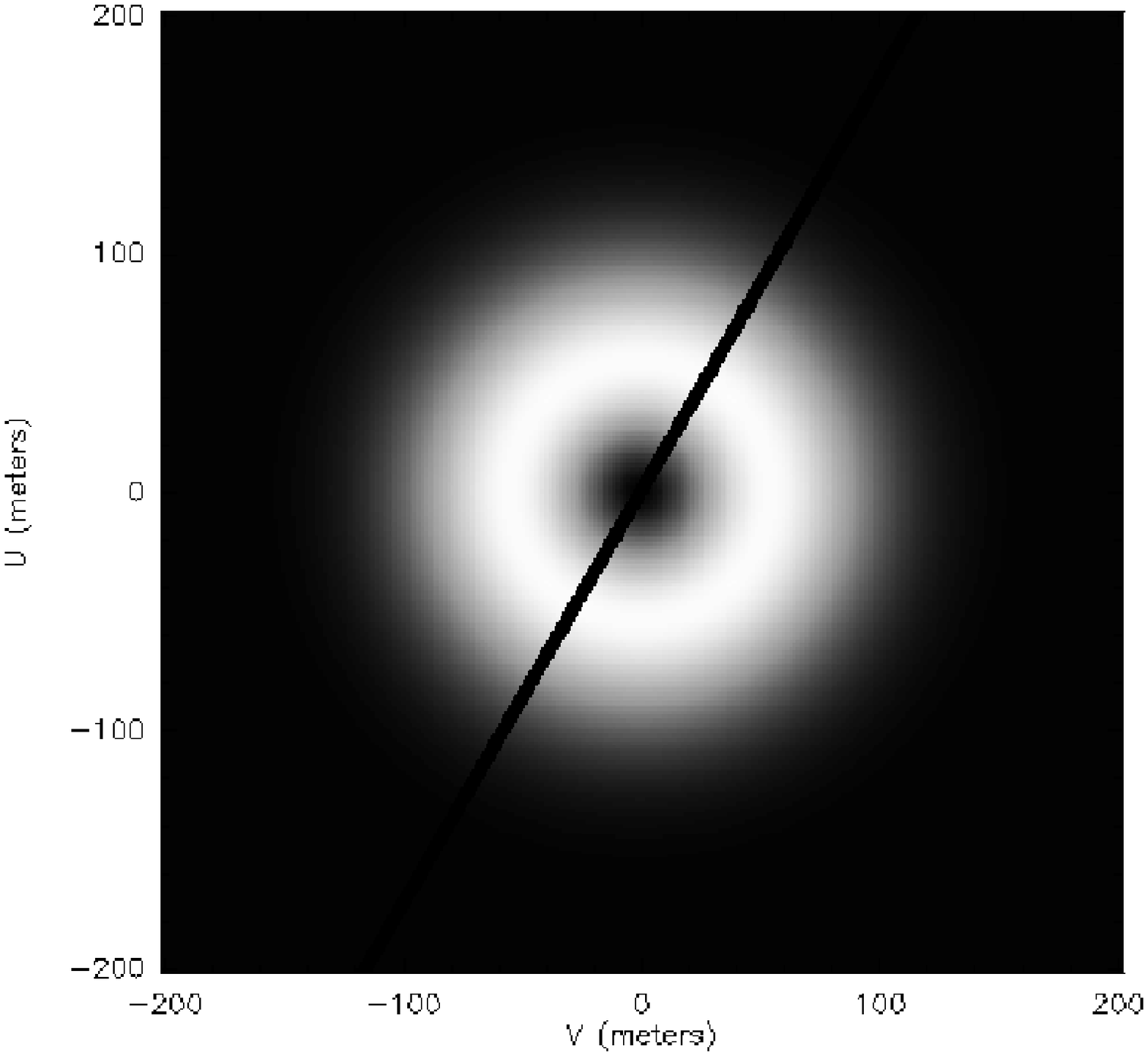}\\
    \includegraphics[width=0.47\textwidth, angle=0]{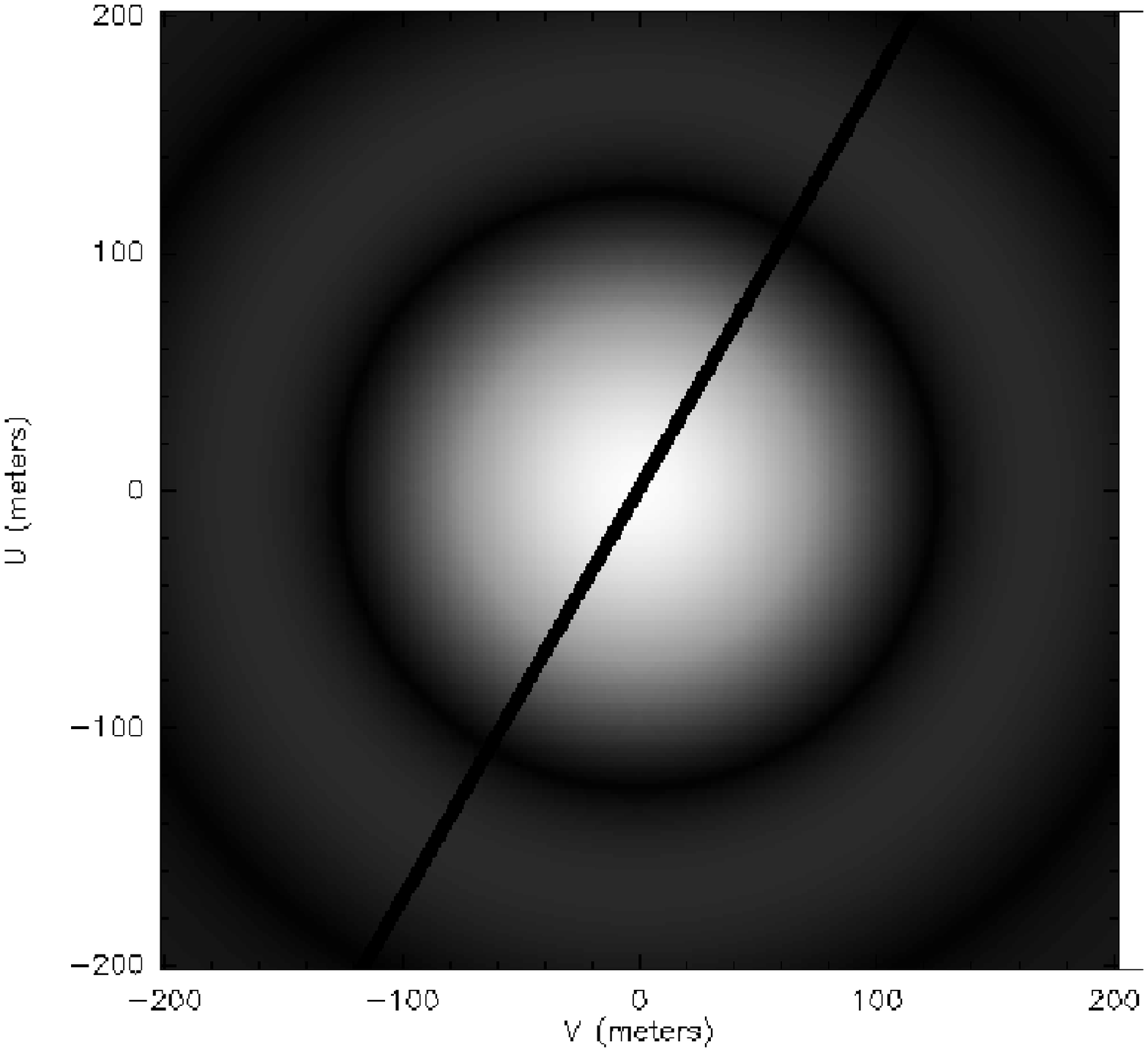}&
    \includegraphics[width=0.47\textwidth, angle=0]{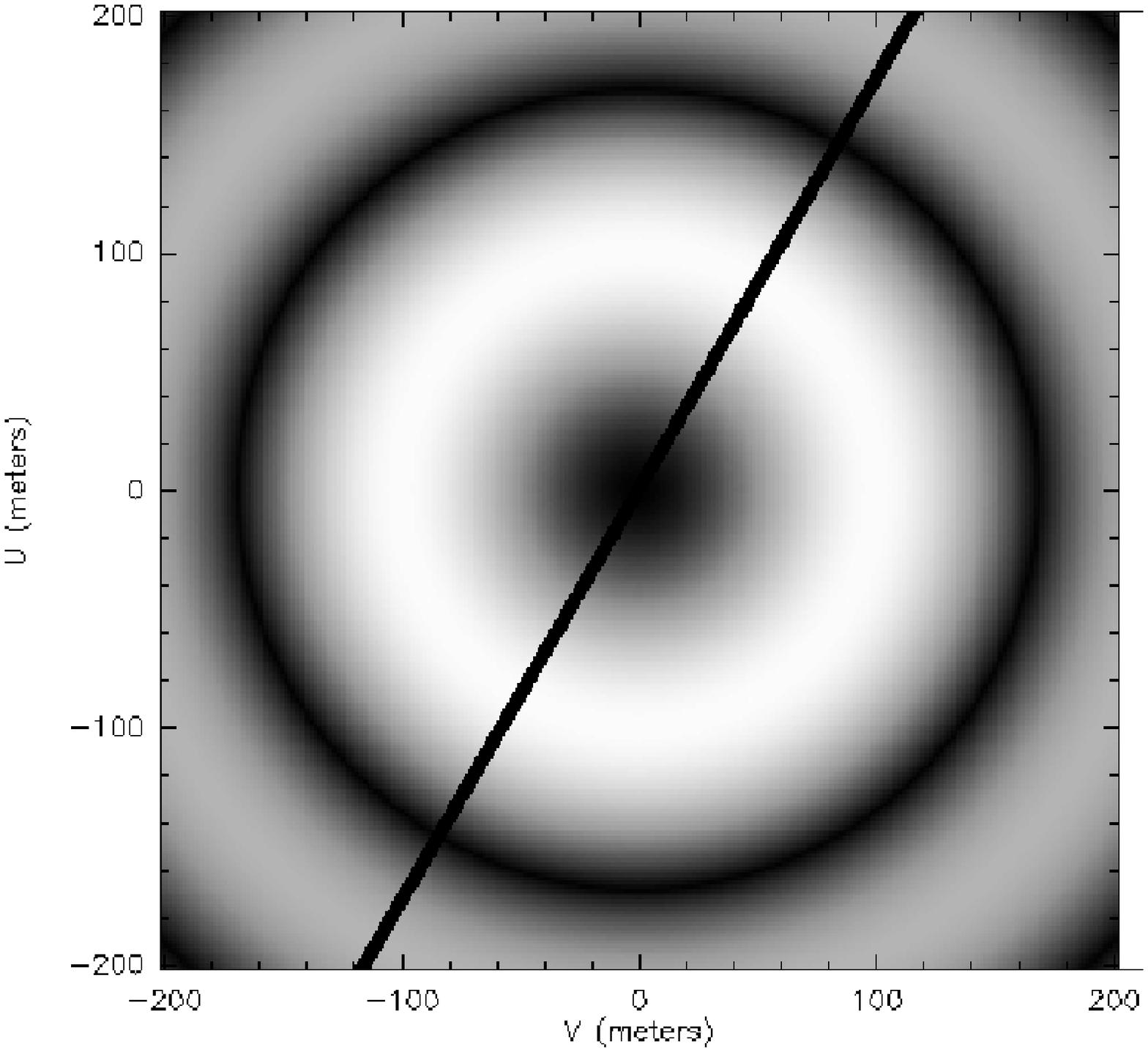}
  \end{tabular}
  \caption[Visibility change versus baseline.]{
    \footnotesize{
      \emph{\bf Top-left:} Gaussian disk visibility map, as a function
      of $UV$ coordinates.
      \emph{\bf Top-right:} Visibility derivative versus the model
      size (FWHM), showing where the visibility varies more with $UV$
      coordinates. 
      \emph{\bf Bottom-left:} Visibility map, but using a uniform disk.
      \emph{\bf Bottom-right:} Visibility derivative versus model size
      (FWHM) for a uniform disk model.
    }
  }
  \label{fig:visiDerivative}
\end{figure}

Fig.~\ref{fig:visiDerivative} shows the model visibility and its
derivative, relative to the model size, for a Gaussian disk and a
uniform disk. 

\paragraph*{Uniform disk:}
The size used is 2\,mas. The derivative peaks where the
visibility slope versus baseline is the largest. This means that for a
small baseline change, a large visibility change will be observed;
i.e., the biggest constraint will be applied to the corresponding
model in the model fitting process.

One can see that the optimal baseline is 100\,m for the uniform
disk. One should know that the optimal baseline to constrain a
given model corresponds to a visibility of about 50\%. Very
important to know: baselines that are too short or too long will not
reveal much information about the source size or shape.

\paragraph*{Gaussian disk:}
The size used is the same as before: 2\,mas. The optimal baseline is
about 50\,m for the Gaussian disk. Therefore, using different baselines,
one will be able to both disentangle the model shape (Sharp - UD - or
smooth - Gauss - edges?) and constrain the typical size (as
seen before).

\subsection*{\underline{Exercise 6:} An unknown astrophysical object.}

\begin{figure}[htbp]
  \centering
  \begin{tabular}{cc}
    \multicolumn{2}{c}{\includegraphics[width=0.3\textwidth, angle=0]{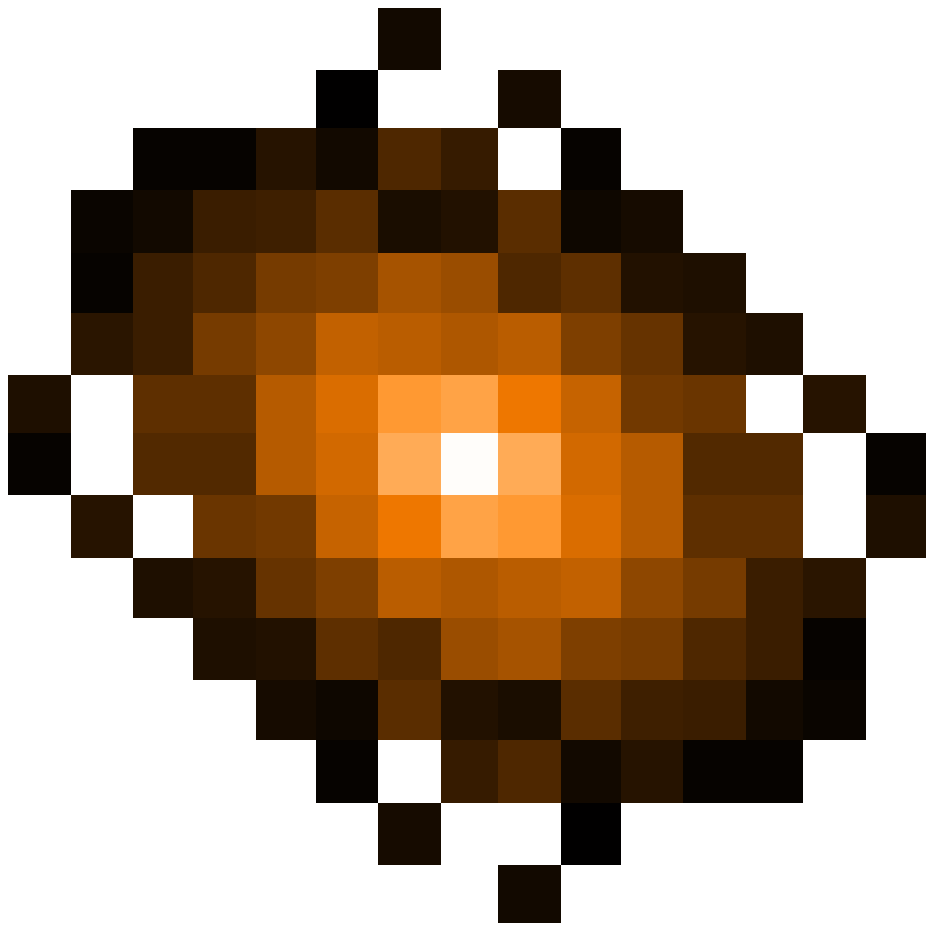}}\\
    \includegraphics[width=0.47\textwidth, angle=0]{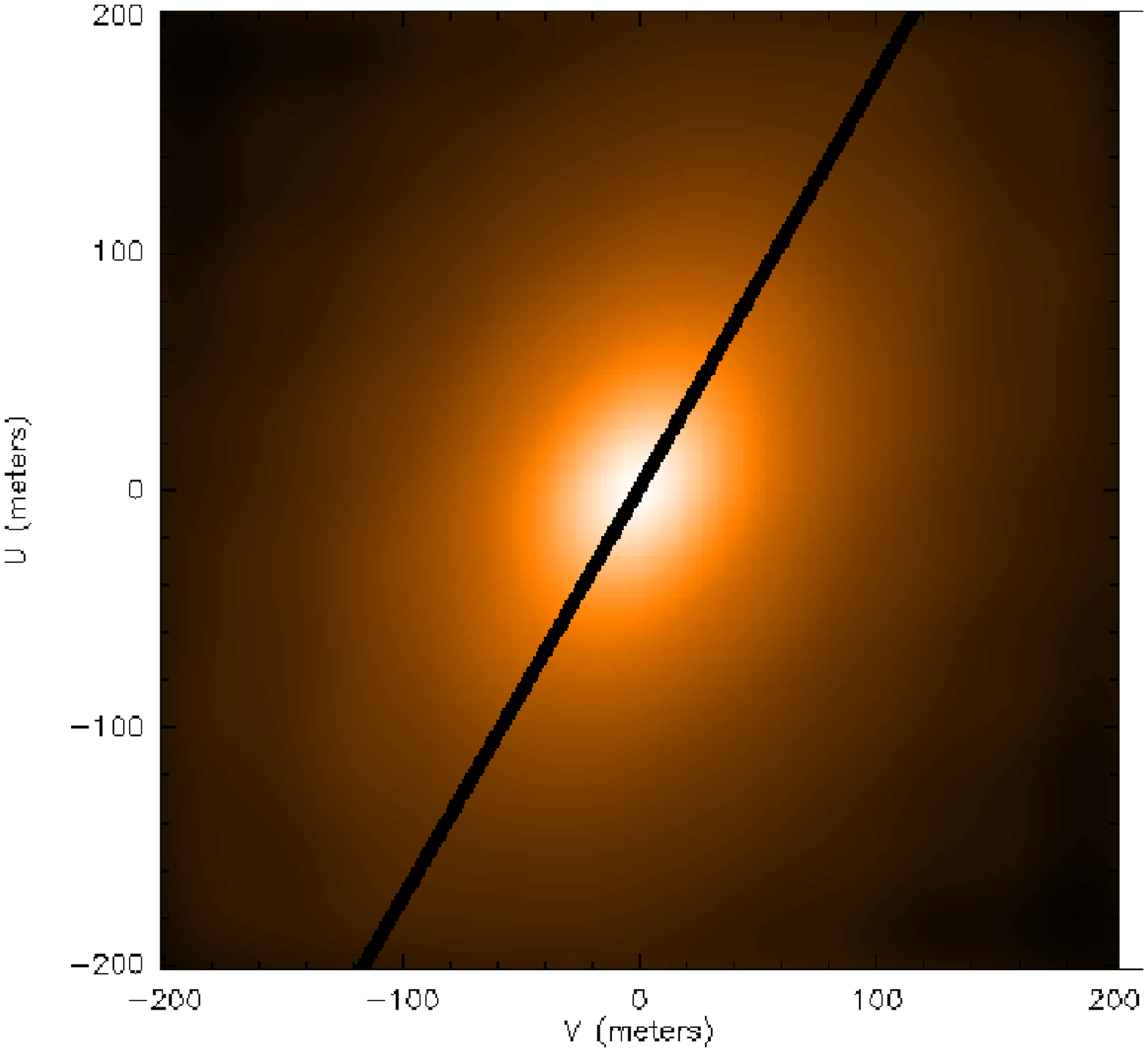}&
    \includegraphics[width=0.47\textwidth, angle=0]{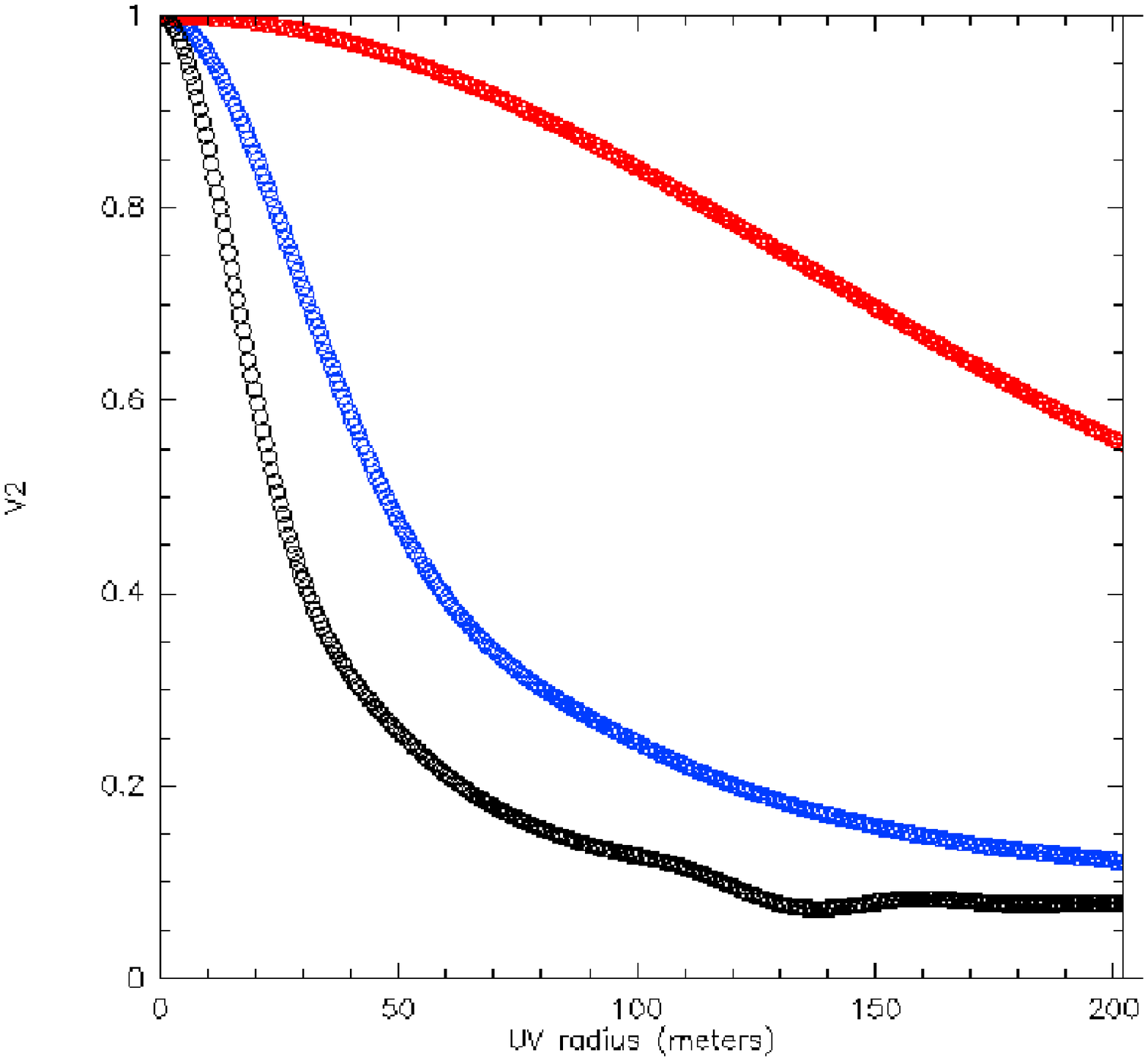}
  \end{tabular}
  \caption[Fu Ori disk model.]{
    \footnotesize{
      \emph{\bf Top:} Image of the disk model used in exercise 6
      (shown using the fv tool)
      \emph{\bf Bottom-Left:} Visibility map showing the main elongation
      of the disk, perpendicular to the main elongation of the
      visibility function.
      \emph{\bf Bottom-Right:} Visibility of the disk for different
      wavelength regimes: red (top curve) is N-band, blue (middle
      curve) is K-band, and black (bottom curve) is J-band.
    }
  }
  \label{fig:FuOriDisk}
\end{figure}

\paragraph*{Loading and displaying a home-made model:}
Fig.~\ref{fig:FuOriDisk} (top) shows the  model of a young stellar disk
produced by F. Malbet. The star-to-disk contrast here is 1 to 10. 

Please note that, as for the previous disk model (Exercise 3), the
disk is elongated, and therefore, the large visibilities will
correspond to the minor axis of the model, whereas the low visibilities
will correspond to the major axis. 

\paragraph*{Computing the visibilities of a home-made model:}
The lower-left part of Fig.~\ref{fig:FuOriDisk} displays a $UV$ map of
the model shown in the upper part. This map is very similar to the one
shown in Fig.~\ref{fig:diskModel}, which indicates that the disk
mostly looks like a Gaussian disk.

\paragraph*{Comparing visibilities for different wavelengths:}
The lower-right part of Fig.~\ref{fig:FuOriDisk} shows the visibility
for the minor axis with different wavelengths: J (black, lower curve),
K (blue, middle curve), and N (red, upper curve). This shows which
wavelength to use to observe the object: the K-band allows both
high visibilities for the extended component and low visibilities for
the detailed structure of the disk, in the range of the VLTI offered
baselines (16-130\,m).

\subsection*{\underline{Exercise 7:} Play with spectral variations, closure phases,
  etc.}

\begin{figure}[htbp]
  \centering
  \begin{tabular}{cc}
    \includegraphics[width=0.47\textwidth]{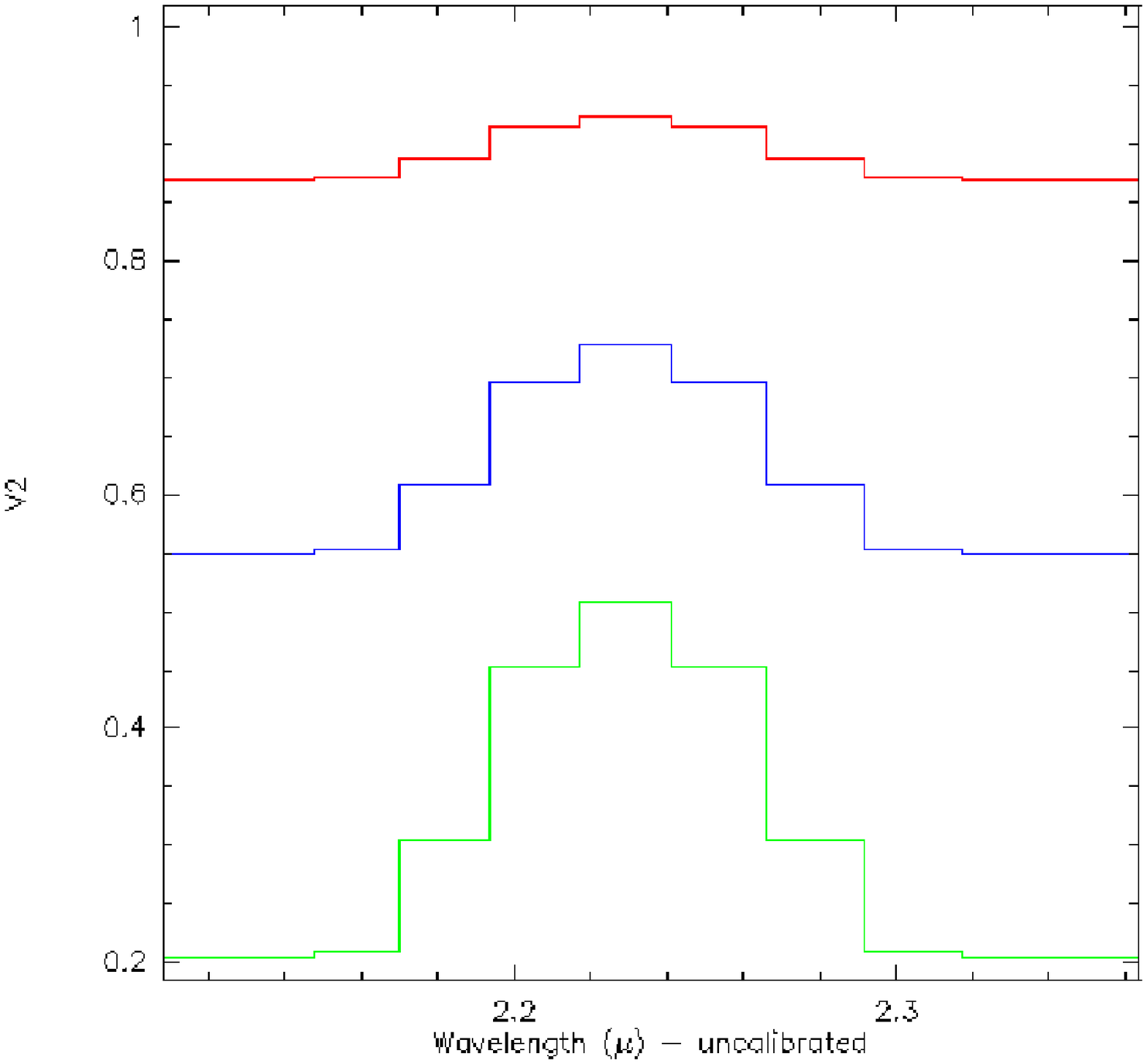}&
    \includegraphics[width=0.47\textwidth]{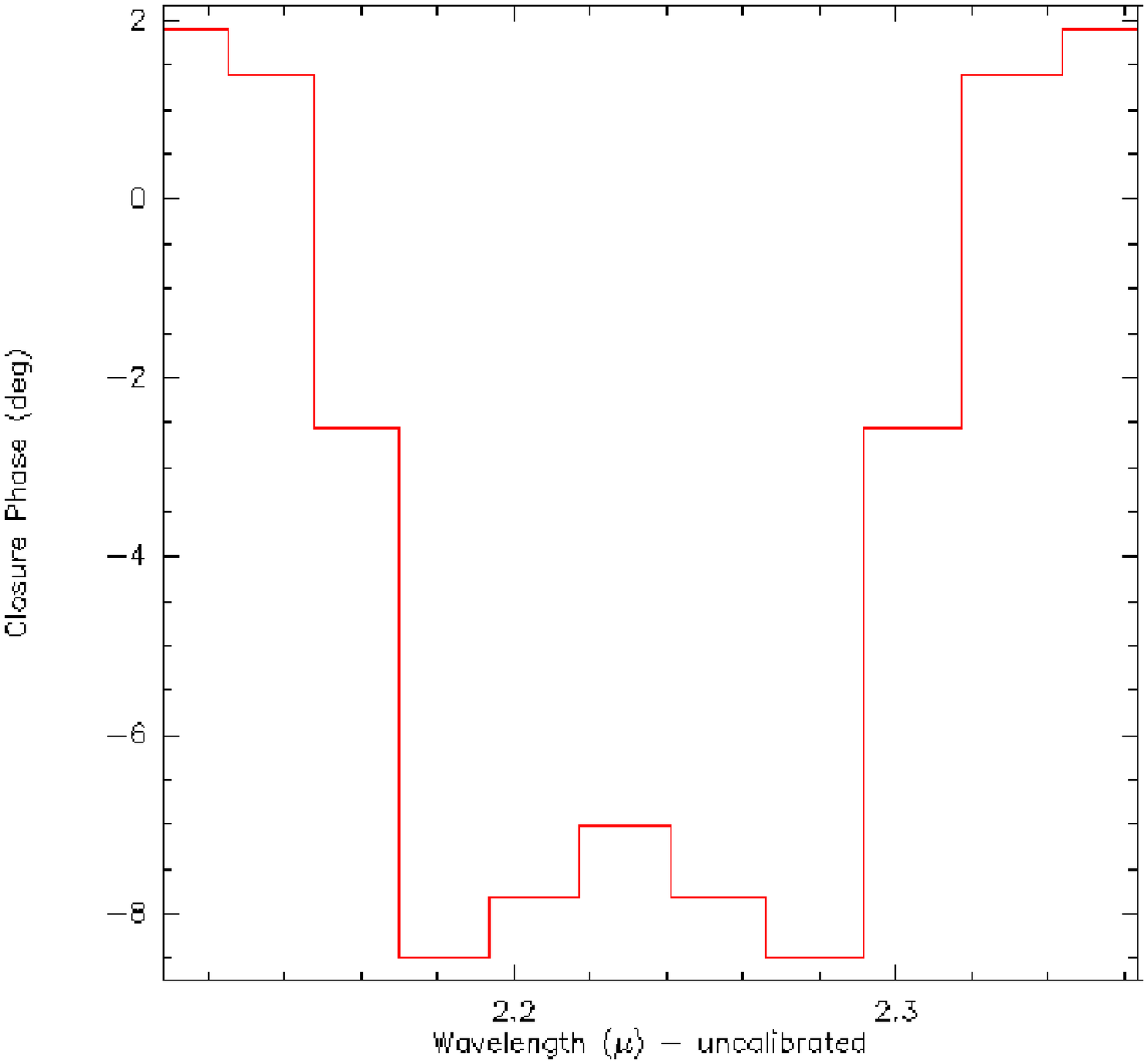}
  \end{tabular}
  \caption[$\gamma^2$ Vel model.]{
    \footnotesize{
      \emph{\bf Left:} Visibility plot as a function of wavelength for
      different baselines (in red, green and blue). The global
      geometry of the model does not change, only the flux ratios
      between the 3 components.
      \emph{\bf Right:} Closure phase plot as a function of
      wavelength.
    }
  }
  \label{fig:gammaVelModel}
\end{figure}

This is a bonus exercise. To manage it, one will need the newest local
version of \texttt{ASPRO} to cope with closure phases (at the time of
writing this correction, the web version of \texttt{ASPRO} did not have all
functionalities necessary to do this exercise). Therefore, one is not
obliged to reach this point.

\paragraph*{Plotting visibilities and closure phase versus wavelength:}
In Fig.~\ref{fig:gammaVelModel}, one can see the visibilities for
a given observational setup (A0-D0-H0) and the closure phase. The
author has used, on purpose, an array with aligned baselines to
illustrate this exercise.

\paragraph*{Qualitative understanding:}

There are a number of indicative clues to qualitatively understand
the shape of the observed object:
\begin{itemize}
\item The different visibilities are decreasing with
  baseline. Therefore, the object is barely resolved by the
  interferometer. One cannot qualitatively distinguish between a
  uniform disk, a Gaussian disk, or a binary star, but one can say the
  object is resolved at the largest baseline ($\approx$130m) and
  therefore has a size of about 2mas.
\item The non-zero closure phase gives information about the
  asymmetry of the object. Here, one has a non-zero but very small
  closure phase. The only simple model  known from this practice
  session which gives a non-zero closure phase is a binary star
  model. The fact that the closure phase is not 180 degrees gives the
  additional information that the flux ratio is not 1/1.
\end{itemize}

Therefore, only qualitatively looking at these data one can say:
\begin{itemize}
\item The object is likely to be a binary star,
\item the separation is about 2mas,
\item the flux ratio is not 1/1,
\item so far, one cannot say if each component has been resolved or if
  there is a third component.
\end{itemize}

\begin{figure}[htbp]
  \centering
  \includegraphics[width=0.4\textwidth]{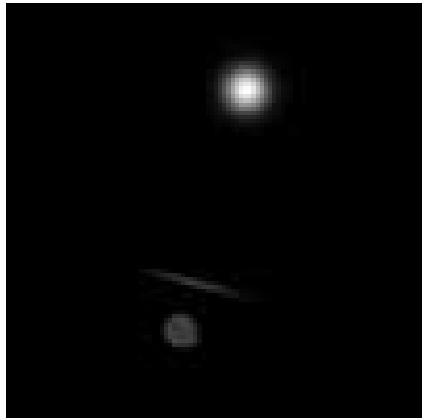}
  \caption[$\gamma^2$ Vel solution.]{
    \footnotesize{
      Image of the binary+wind-wind collision zone model used in this
      exercise.
    }
  }
  \label{fig:gammaVelModel2}
\end{figure}

\paragraph*{Looking at the solution:}
Fig.~\ref{fig:gammaVelModel2} shows the image of the model used in
this exercise. As one can see, the components are somewhat resolved,
but probably not enough to be detected, and a third component is
present. Therefore, the qualitative analysis is not enough, and one
has to perform a quantitative analysis to characterize all of these
components.

\newpage

\section{Observability and $UV$ coverage}

\subsection*{What you will need for this particular practice session:}

You will need to use \texttt{ASPRO} and the catalogs named
\texttt{sampleSources1.sou} and \texttt{sampleSources2.sou} provided with
this practice session (you can copy this information into text files,
see below). To set up \texttt{ASPRO}, please refer to the 1st part of
this practice session about visibilities and model fitting.

\begin{verbatim}
!-----------------------------------------------------
! sampleSources1.sou
!-----------------------------------------------------
ACHERNAR    EQ 2000.000    01:37:42.8466 -57:14:12.327
NGC_1068    EQ 2000.000    02:42:40.8300 -00:00:48.400
BETELGEUSE  EQ 2000.000    05:55:10.3053 +07:24:25.426
HD_68273    EQ 2000.000    08:09:31.9503 -47:20:11.716
HD_81720    EQ 2000.000    09:25:19.2802 -54:27:49.559
\end{verbatim}

\begin{verbatim}
!----------------------------------------------
! sampleSources2.sou
!----------------------------------------------
STAR_40   EQ 2000.000    16:00:0.0    40:00:0.0
STAR_30   EQ 2000.000    16:00:0.0    30:00:0.0
STAR_20   EQ 2000.000    16:00:0.0    20:00:0.0
STAR_10   EQ 2000.000    16:00:0.0    10:00:0.0
STAR_0    EQ 2000.000    16:00:0.0     0:00:0.0
STAR_-20  EQ 2000.000    16:00:0.0   -20:00:0.0
STAR_-40  EQ 2000.000    16:00:0.0   -40:00:0.0
STAR_-60  EQ 2000.000    16:00:0.0   -60:00:0.0
STAR_-80  EQ 2000.000    16:00:0.0   -80:00:0.0
\end{verbatim}

\subsection*{\underline{Exercise 1:} Setting up an observation}

\paragraph*{Set the date:} In the \emph{WHEN} menu, {\it Date \& Time
  Setup}, put the date \texttt{28-AUG-2007} and time
\texttt{14:00:00}.
\paragraph*{Set the place:} In the \emph{WHERE} menu, select {\it VLT,
  2 Telescopes}.
\paragraph*{Set the target:} In the \emph{WHAT} menu, choose {\it Use
  Object catalog} and select the file \texttt{sampleSources1.sou}. If
you use the web version of \texttt{ASPRO}, you need to have an account on the
JMMC server and to copy your files beforehand using the {\it File
  Management} panel.
\paragraph*{Check the settings:} Check with the \emph{WHAT} menu
{\it View Object catalog} and look at the result in the xterm
window. Note: you can do this step only with a local version of
\texttt{ASPRO} as the web version has no access to a terminal.

\subsection*{\underline{Exercise 2:} Observability of sources at different
  declinations and delay line constraints}

First, we will check the observability of the sources with {\it
  OBSERVABILITY/COVERAGE, Observability of Source}. Set the minimum
elevation to 30\deg, check the {\it Plot the twilight zones}
button, and use UT1 and UT2.  When everything is done, press the {\it GO}
button.

\paragraph*{Which stars are observable?} Is the chosen date appropriate for
observing all  stars together?
\paragraph*{Delay lines limitation:} Now, go to {\it
  OBSERVABILITY/COVERAGE, Observability limits due to delay
  lines}. How does it change the observability? Compare the
observability with UT1-UT4 and G1-J6. What do you conclude?

\subsection*{Sampling the $UV$ plane with the VLTI}

This goal of this section is to see how the $UV$ coverage changes with
baseline orientation and source declination.

You should first load the catalog named
\texttt{sampleSources2.sou}. It contains 7 stars of R.A.  5:00:00 and
of different declinations. In this section, you will make
intensive use of the \emph{OBSERVABILITY/UV COVERAGE} menu of
\texttt{ASPRO}.

\begin{figure*}[htbp]
  \begin{center}
    \includegraphics[height=1.0\hsize, angle=0]{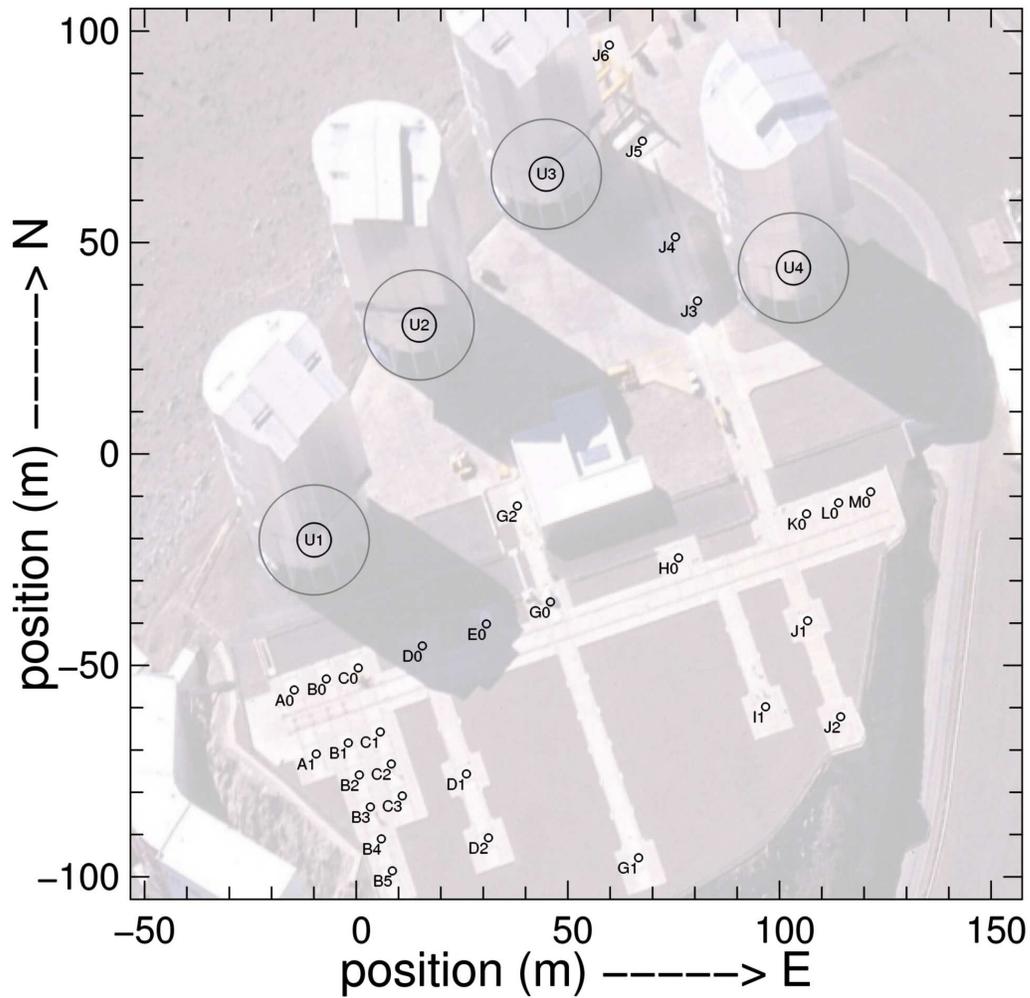}
  \end{center}
  \caption{The VLTI stations. Photo: Gerhard H\"udepohl}
  \label{fig3}
\end{figure*}

\subsection*{\underline{Exercise 3:} $UV$ tracks for a North-South baseline}

\begin{figure*}[htbp]
  \begin{center}
    \begin{tabular}{cc}
      \includegraphics[angle=0,width=0.45\hsize]{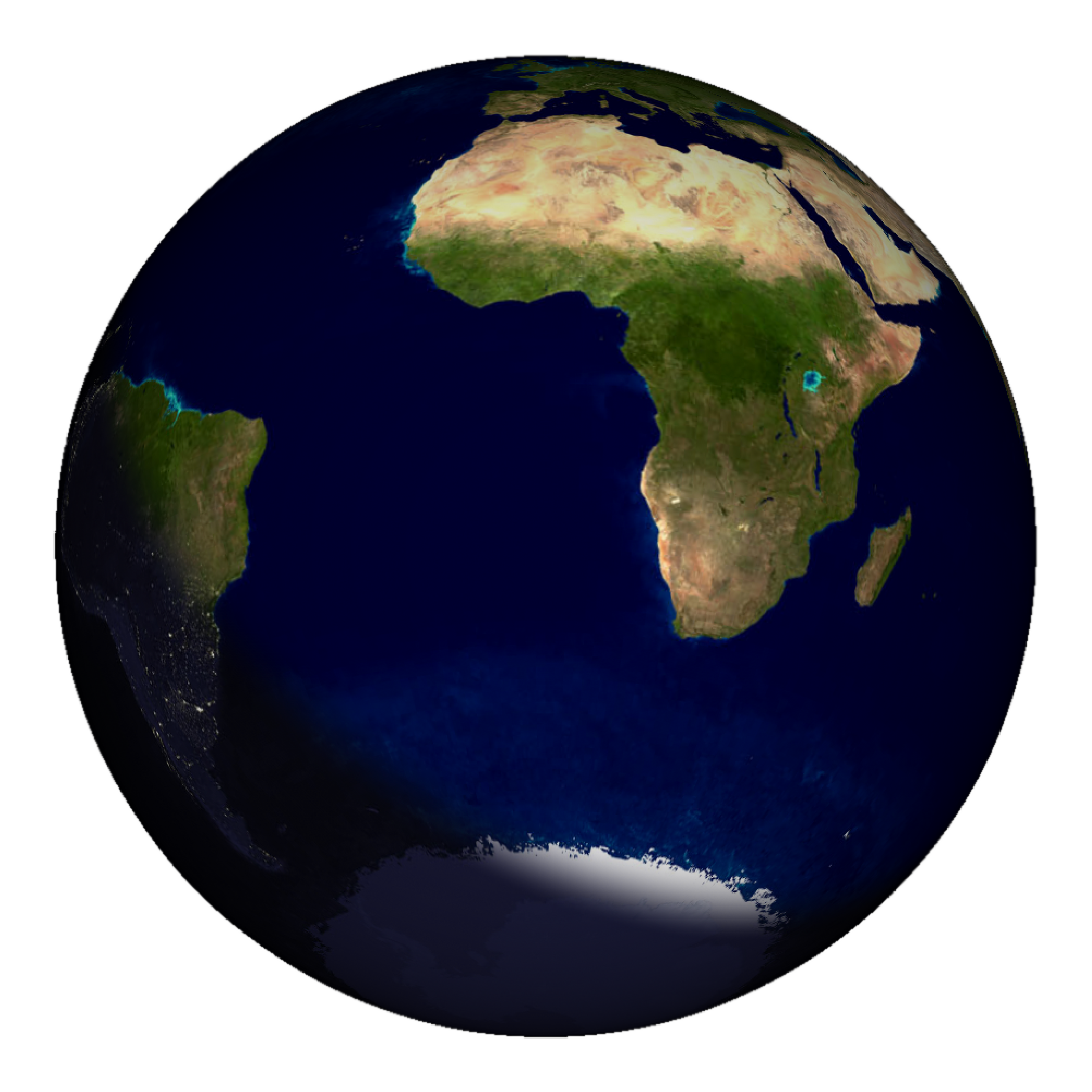}&
      \includegraphics[angle=0,width=0.45\hsize]{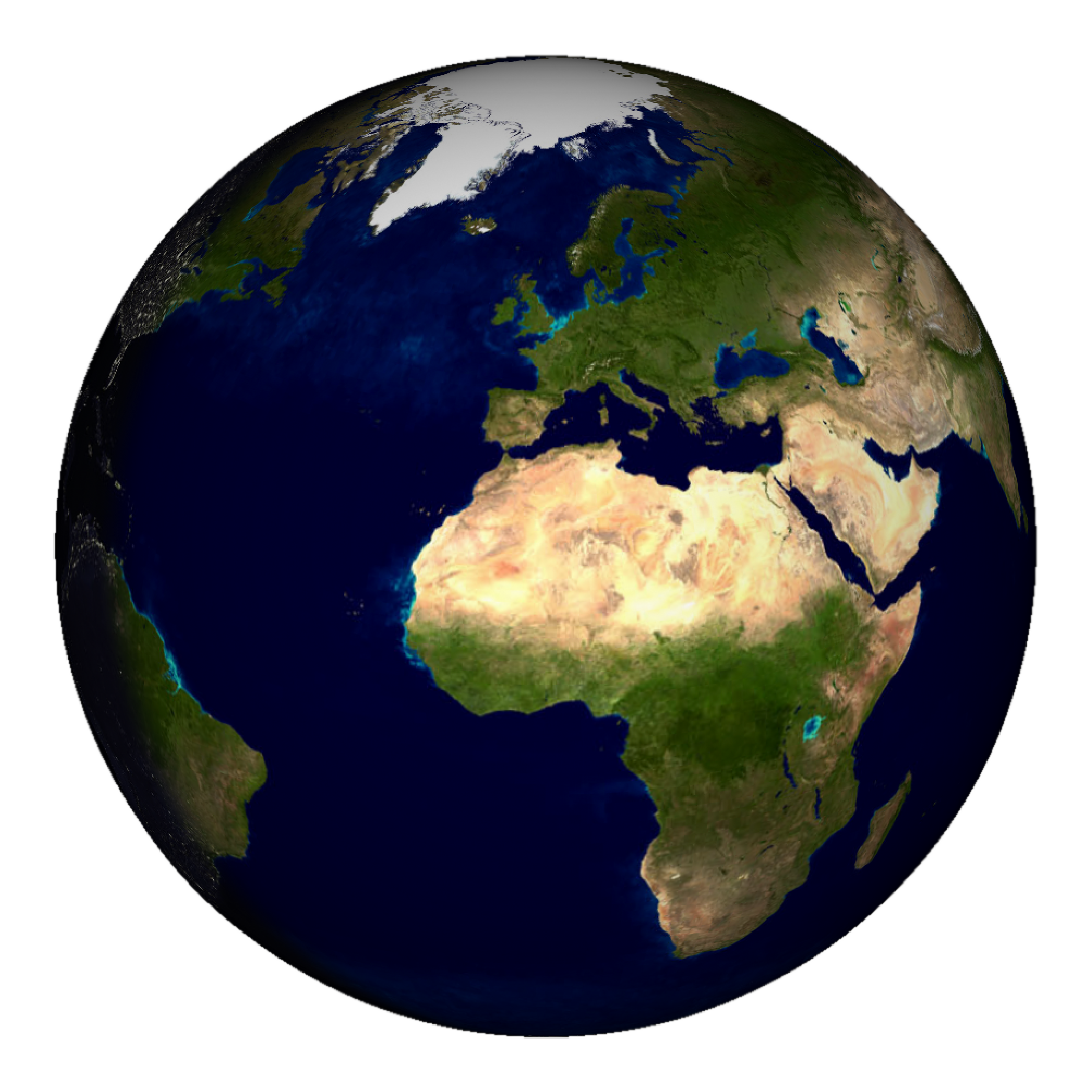}
      \footnote{}
    \end{tabular}
  \end{center}
  \caption{
    The Earth, as seen from a -25\deg\ declination (left) and a
    +25\deg declination (right). These images were generated
    using the free software xplanet,
    \url{http://xplanet.sourceforge.net}.
  }
  \label{fig1}
\end{figure*}

We will now study the $UV$ coverage of the sources with {\it
  OBSERVABILITY/COVERAGE, UV coverage \& PSF}. 

\begin{itemize}
\item Select the star at declination -20 and set the wavelength to 2
  microns. In the {\it Telescopes \& Stations} panel, select a
  2-telescope baseline oriented N-S (cf.\ Fig.\ \ref{fig3} ) and have
  a look to the shape of $UV$ coverage you get.
\item Change stars, going from positive to negative declination
  and see what happens (you can over-plot the graphs by unchecking
  the \emph{RESET FRAME} button in the {\it Telescopes \& Stations}).
\end{itemize}

\emph{Hint: Look at the orientation of the Earth in Fig.\ref{fig1}.}

\subsection*{\underline{Exercise 4:} $UV$ tracks for an East-West baseline}

Select a large 2-telescope baseline oriented E-W.  Visualize the
observability of the targets and check the delay line constraints.
Plot the $UV$ coverage for several stars. 

\paragraph*{$UV$ tracks shape:} Why are the $UV$-tracks elliptical (you
can refer to the interferometry introduction articles)?

\paragraph*{$UV$ tracks and target declination:} Have a look at the
$UV$-tracks of a star above the equator and below the equator. What do
you notice?

\emph{Hint : Look at the figure of the Earth in Fig.\ref{fig1} again.} 
\bigskip

\paragraph*{Importance of the baseline orientation:} Compare the N-S
baseline and the E-W baseline in terms of $UV$-coverage and
observability (how much $UV$-track do you cover with the same fixed
delay?) Play with the star and the end of hour angle range.

\subsection*{\underline{Exercise 5:} $UV$ tracks for a 3-telescope-array}

\paragraph*{Observability:}
\begin{itemize}
\item Select a large 3-telescope array configuration (in the
  \emph{WHERE} menu).
\item Visualize the observability of the targets (including
  constraints on delay lines). Look at the {\it
    OBSERVABILITY/COVERAGE, Observability limits due to delay lines}
  panel to see why the observability range is smaller with 3
  telescopes than with 2 telescopes.
\end{itemize}

\paragraph*{$UV$ tracks for different configurations:} In the \emph{UV
  coverage} panel, try to add several 3-telescope configuration. For
that you need to un-check the {\it reset frame} button. As an example,
you can select 4 configurations i.e.\ \texttt{A0-G1-J6},
\texttt{G2-J1-D2} and \texttt{A1-B2-C1}.

\paragraph*{Beam shape:} You can then display the ``dirty beam'' (the
same as in radio-astronomy !) by using the {\it Display PSF} panel.




\subsection*{\underline{Exercise 6:} Radius measurement of a star (uniform disk)}

Here you will play with configurations and ``real'' observations. You
will have a set of stars you want to observe. You must figure out if
they are observable and choose the best observing setup to accurately
measure the diameters.

In this part, you should load the catalog named
\texttt{sampleSources1.sou}. Select an instrument and the K band
(2.1$\mu$m). You should also select an observing period and an optimal
array configuration to determine the radius of the targets with the
highest accessible accuracy. In this section you will make intensive
use of the \emph{WHAT \& Object Model} menu (or \emph{UV Model/FIT,
  Source modeling} menu) and \emph{OBSERVABILITY/COVERAGE} menu of
\texttt{ASPRO}.

\begin{table*}[htbp]
  \centering
  \caption{
    \footnotesize{
      Star main characteristics of catalog \texttt{sampleSources2.sou}
    }
  }
  \label{tab1}
  \medskip
  \begin{tabular}{|c|c|c|c|c|c|c|}
    \hline
    Object	& Spectral Type & Ra & Dec & Diameter\\
    & from simbad & & & [mas] \\
    \hline
    Betelgeuse	& M2Iab	& 05:55:10.31 & +07:24:25.4 & 44.20\\
    Achernar	& B3Ve	& 01:37:42.85 & -57:14:12.3 & 2.53\\
    HD 81720	& K2III	& 09:25:19.28 & -54:27:49.6 & 0.93\\
    HD 68273	& K2III	& 08:09:31.95 & -47:20:11.7 & 0.5\\
    NGC 1068    & AGN   & 02:42:40.83 & -00:00:48.4 & 3 \\
    \hline
  \end{tabular}
\end{table*}

Use the appropriate uniform disk model to either display the
amplitude, the phase of the visibility, or the derivatives with respect
to the diameter to visualize which part of the $UV$ plane really
constrains the model.

\paragraph*{Optimizing the observability of a series of sources:}
Can you find a setup which fits well all the stars together? For that
purpose, you must find a night and configuration which fits well all the
stars characteristics for observability, visibility level, delay lines
constraints, and $UV$ tracks.

\paragraph*{Radius measurement:} Can one determine the radius of these
stars?

\paragraph*{More details about the object:} Can one determine
phenomena that occur at higher spatial frequencies, like limb
darkening? 

\paragraph*{Knowing the limitations:} What accuracy do you need to
fulfill your objectives?

\subsection{\underline{Exercise 7:} Binary parameter determination}

In fact a ``mistake'' was introduced in the previous list: the star
HD~68273 is a binary star (real name $\gamma^2$ Velorum). First
load/re-load the catalog named \texttt{sampleSources1.sou} and then
select star HD~68273. Let us consider it as a binary system with
the properties summarized in Table~\ref{tab2}. 

\begin{table*}[htbp]
  \centering
  \caption{
    \footnotesize{
      Binary system characteristics
    }
  }
  \label{tab2}
  \medskip
  \begin{tabular}{|c|c|c|c|c|}
    \hline
    Ra     &Dec      &$\rho$ &P.A.   & $\Delta$mag\\
    &         &[mas]  &[\deg.] &            \\
    \hline     08:09:31.9503 & -47:20:11.716 & 3.65 & 75 & 0\\
    \hline
  \end{tabular}
\end{table*}

\begin{itemize}
\item Select the baseline G2-G1
\item Visualize the $UV$ coverage and the amplitude.
  Does this baseline constrain the parameters of the binary?
  Plot the visibility as a function of time.
\item Select the baselines A0-M0. Visualize the amplitude, the phase,
  and their derivatives.
\item Does this baseline constrain the parameters of the binary?
\item Plot the visibility as a function of time.
\item What do you notice about the baseline orientation / the binary
  system position angle? 
\end{itemize}

\newpage

\section{Observability and $UV$ coverage (Correction)}

The same procedure as before was used to produce these corrections:
\texttt{ASPRO} was used for getting the figures and \texttt{GIMP} to
over-plot the graphs together for illustration. {\bf As in the
  previous section, please try the exercises first before reading
  these corrections}.

\subsection*{\underline{Exercise 1:} Setting up an observation}

\begin{figure}[htbp]
  \centering
  \includegraphics[width=0.99\textwidth]{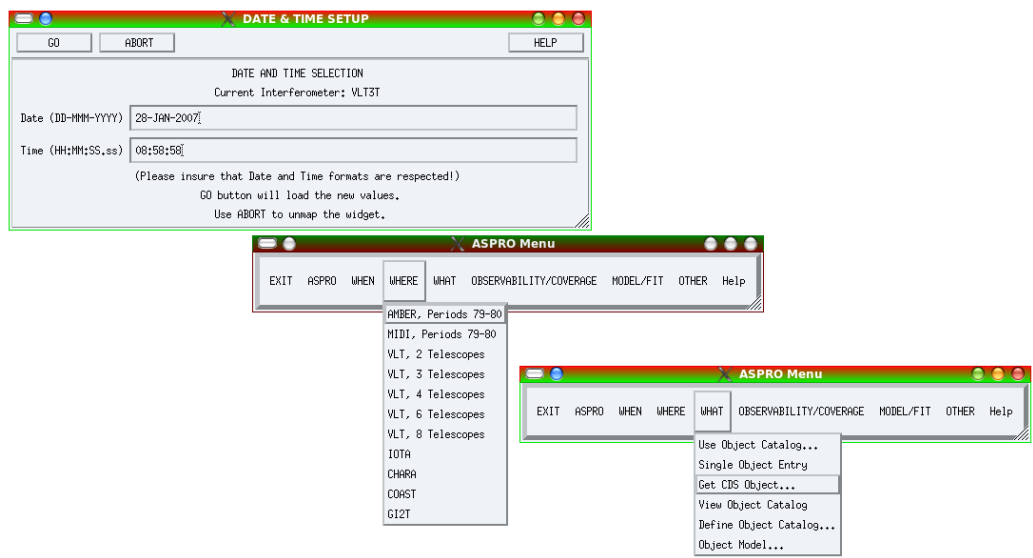}
  \caption[Set date, place and target.]{
    \footnotesize{
      How to set the date, place, and object. Top: the date setup
      window appearing when pressing the \emph{WHEN} menu. After
      changing the date and time, just press \emph{GO}. Middle: the
      \emph{WHERE} menu, where one can choose
      between all the supported interferometers. Bottom: the
      \emph{WHAT} menu. The \emph{Get CDS Object...}  is very
      convenient for quickly finding a target.
    }
  }
  \label{fig:startup}
\end{figure}

After having set the date, time, place, and target
(Fig.~\ref{fig:startup}), you can start using \texttt{ASPRO} and its many
features to check and prepare observations.
Fig.~\ref{fig:catalogPlot} is what you should get in your shell when
setting \emph{View Object Catalog} correctly.

\begin{figure}[htbp]
\begin{verbatim}
Aspro>
!-----------------------------------------------------
! sampleSources1.sou
!-----------------------------------------------------
ACHERNAR    EQ 2000.000    01:37:42.8466 -57:14:12.327
NGC_1068    EQ 2000.000    02:42:40.8300 -00:00:48.400
BETELGEUSE  EQ 2000.000    05:55:10.3053 +07:24:25.426
HD_68273    EQ 2000.000    08:09:31.9503 -47:20:11.716
HD_81720    EQ 2000.000    09:25:19.2802 -54:27:49.559
\end{verbatim}
  \caption[Catalog display.]{
    \footnotesize{
      Display of the catalog in the terminal window.
    }
  }
  \label{fig:catalogPlot}
\end{figure}

\subsection*{\underline{Exercise 2:} Observability of sources at
  different declinations and delay line constraints}

\begin{figure}[htbp]
  \centering
  \begin{tabular}{cc}
    \includegraphics[width=0.47\textwidth]{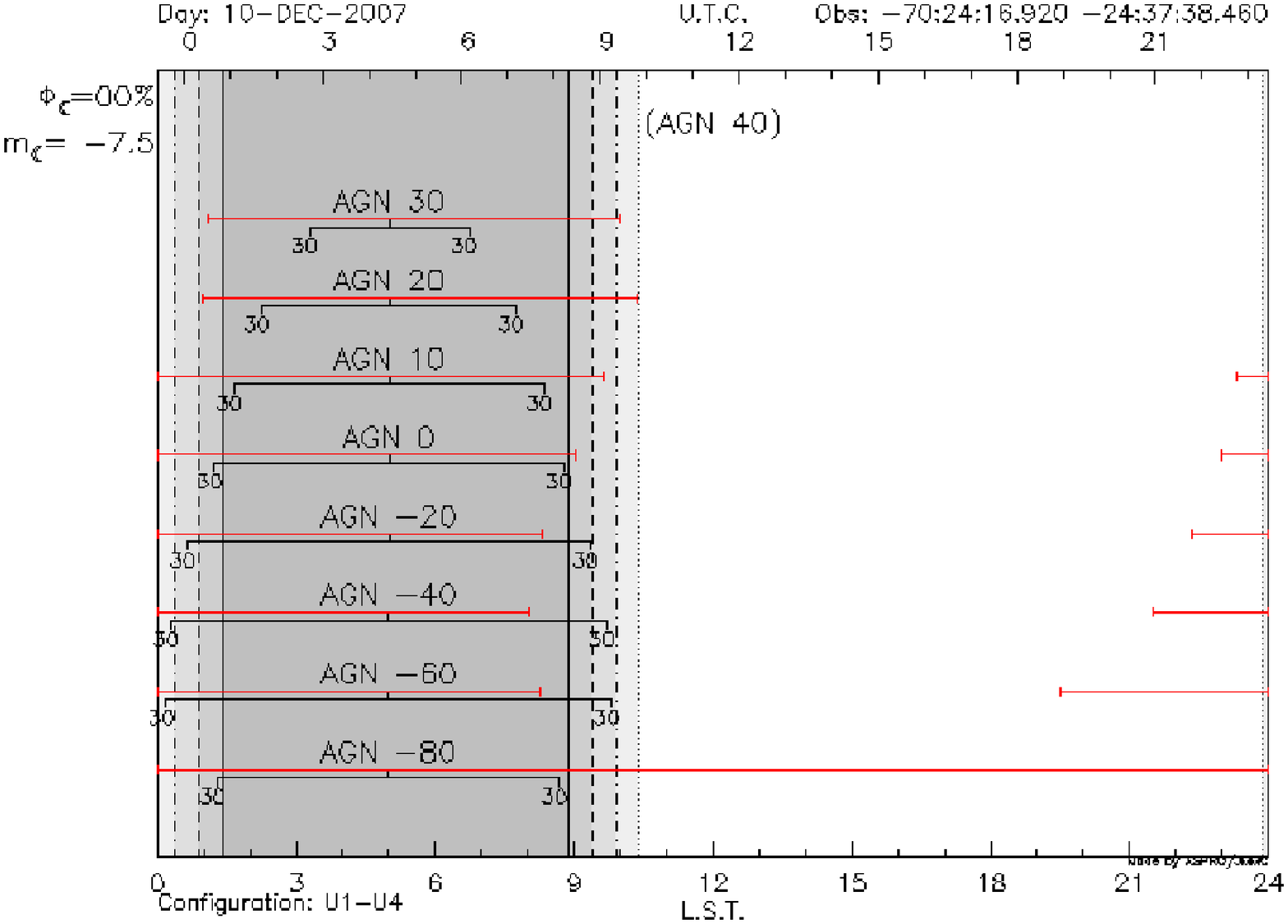}&
    \includegraphics[width=0.47\textwidth]{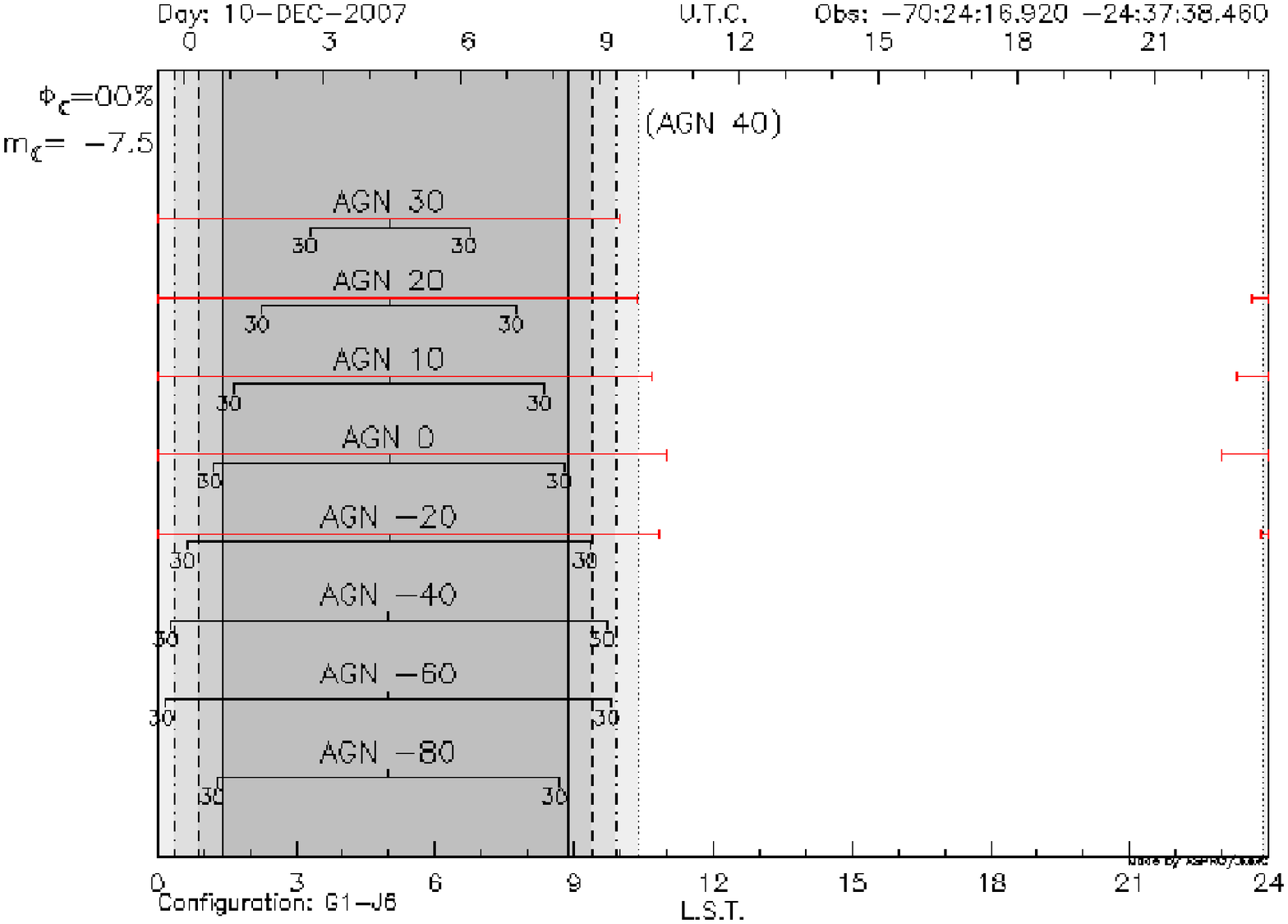}
  \end{tabular}
  \caption[Observability of sources.]{
    \footnotesize{
      \emph{\bf Left:} Observability of sources at different
      declinations for the UT1-UT4 baseline (roughly East-West
      baseline).
      \emph{\bf Right:} Observability of sources for the G1-J6
      baseline (North-South).
    }
  }
  \label{fig:observability}
\end{figure}

\paragraph*{Which stars are observable?} 
Fig.~\ref{fig:observability} shows the observability for the second
catalog of sources (\texttt{sampleSources2.sou}). The gray area
corresponds to the night, and the black lines correspond to the
observability of sources by their height above the horizon (here
30\deg). First, you can see that a source at declination +40 is not
observable at all, due to the latitude of Paranal: about 30\deg.

\paragraph*{Delay lines limitation:} 
The additional constraints coming from the interferometer are
displayed in red just above the usual observability plot: the left
graph is for stations UT1-UT4 and the right one
is for G1-J6. You can see the difference for 2 extreme cases: a
(roughly) East-West baseline (UT1-UT4) and a North-South baseline
(G1-J6). The North-South baseline restricts the access to northern
sources, whereas the East-West baseline restricts the observability
during the night.

\subsection*{\underline{Exercise 3:} $UV$ tracks for a North-South baseline}

\begin{figure}[htbp]
  \centering
  \includegraphics[width=0.7\textwidth]{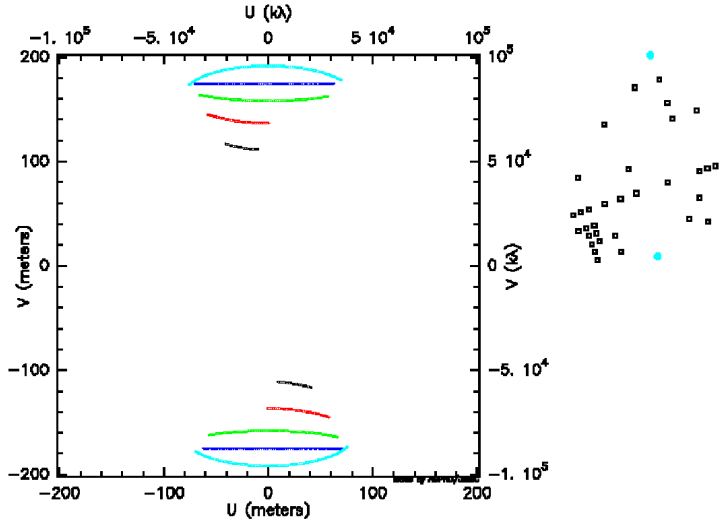}
  \caption[Fu Ori disk model.]{
    \footnotesize{
      $UV$ plane tracks due to the Earth rotation for a N-S baseline
      (G1-J6)  and for different declinations: cyan for -20\deg, blue
      for 0\deg, green for 10\deg, red for 20\deg\ and black for
      30\deg.
    }
  }
  \label{fig:UVplaneNS}
\end{figure}

The $UV$ tracks for the targets of the 2nd catalog
(\texttt{sampleSources2.sou}) are shown in
Fig.~\ref{fig:UVplaneNS}. The North-South baseline gives roughly
North-South $UV$ tracks, but the limitations due to the delay line
limits are quite severe here since a very long baseline (G1-J6) was
used. This can be seen in the very short accessible $UV$ tracks for
the 20 and 30\deg\ declination targets. One can see here that the
0\deg\ declination target gives a line in the $UV$ plane.

\subsection*{\underline{Exercise 4:} $UV$ tracks for an East-West baseline}

\paragraph*{$UV$ tracks shape:} For the answer to this question, please
refer to \citet{2008_Millour}. The $UV$ tracks correspond to the
projection of a circle (due to the Earth rotation) on an inclined
plane (the plane of sky) and are therefore arcs of ellipses.

\begin{figure}[!h]
  \centering
  \includegraphics[width=0.7\textwidth]{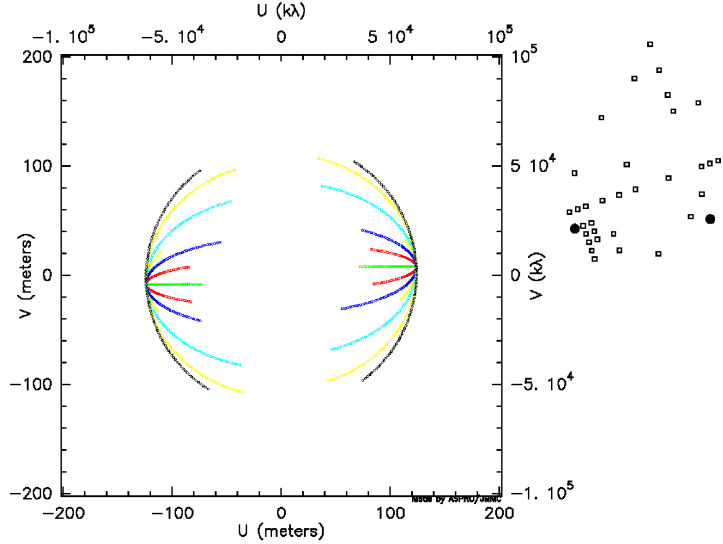}
  \caption[Fu Ori disk model.]{
    \footnotesize{
      $UV$ plane tracks due to the Earth rotation for an E-W baseline
      (A1-J2). The color-coding for the different
      declinations is the following: black for -80\deg, cyan for
      -40\deg, blue for -20\deg, green for 0\deg, red for 20\deg\ and
      yellow for 40\deg.
    }
  }
  \label{fig:UVplaneEW}
\end{figure}

\paragraph*{$UV$ tracks and target declination:} 
Fig.~\ref{fig:UVplaneEW} now shows the East-West baseline. As one can
see, East-West baselines will never give a
North-South $UV$ projection on the sky. Also, the total range of $UV$ plane
projected angles is maximum for high declination targets (either
positive or negative). As for the previous case, a 0\deg\ target gives
a straight line in the $UV$ plane and, therefore, provides less coverage
than a high declination target.

\subsection*{\underline{Exercise 5:} $UV$ tracks for a 3-telescope-array}

\begin{figure}[htbp]
  \centering
  \begin{tabular}{cc}
    \includegraphics[width=0.47\textwidth]{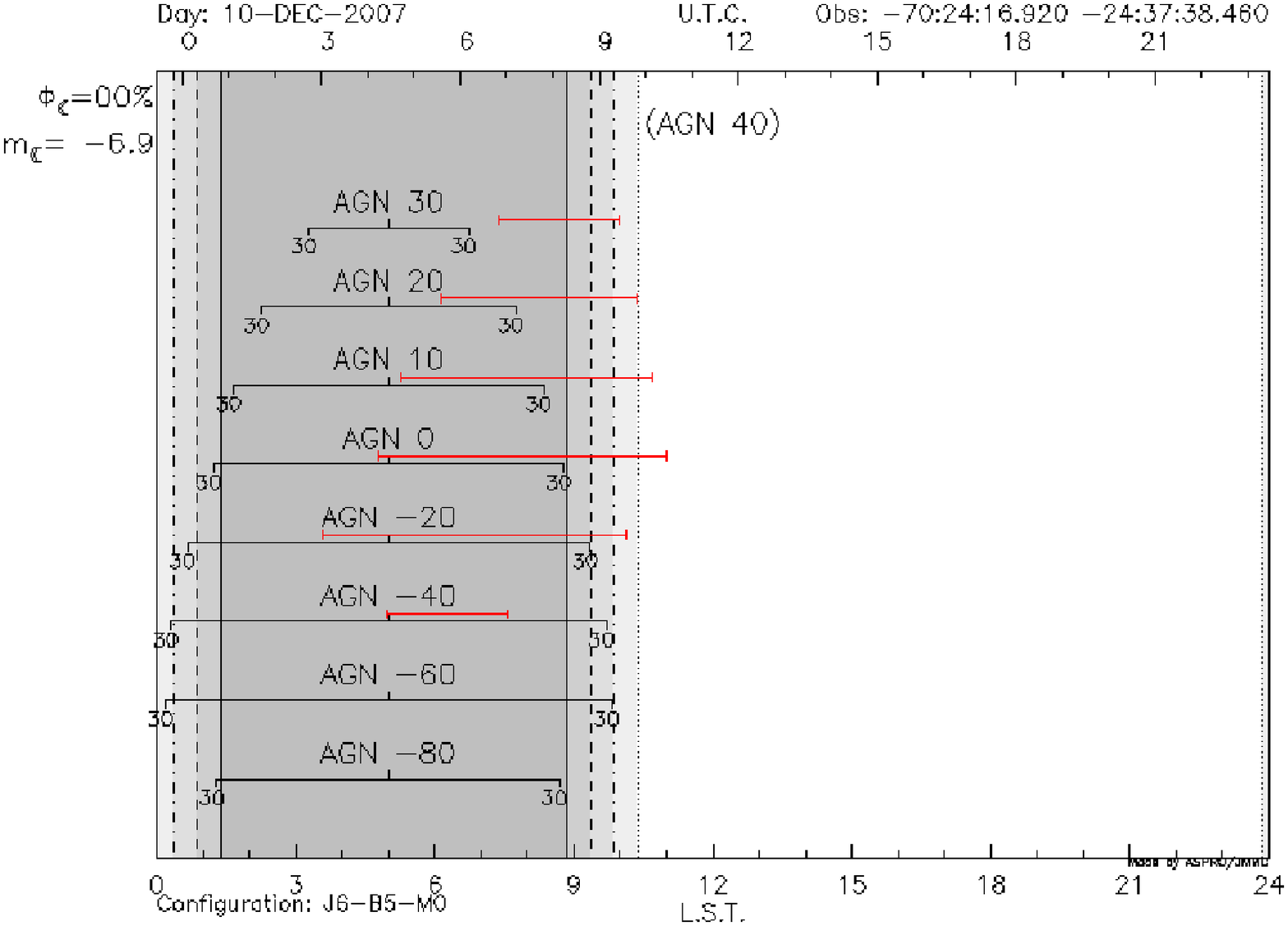}&
    \includegraphics[width=0.47\textwidth]{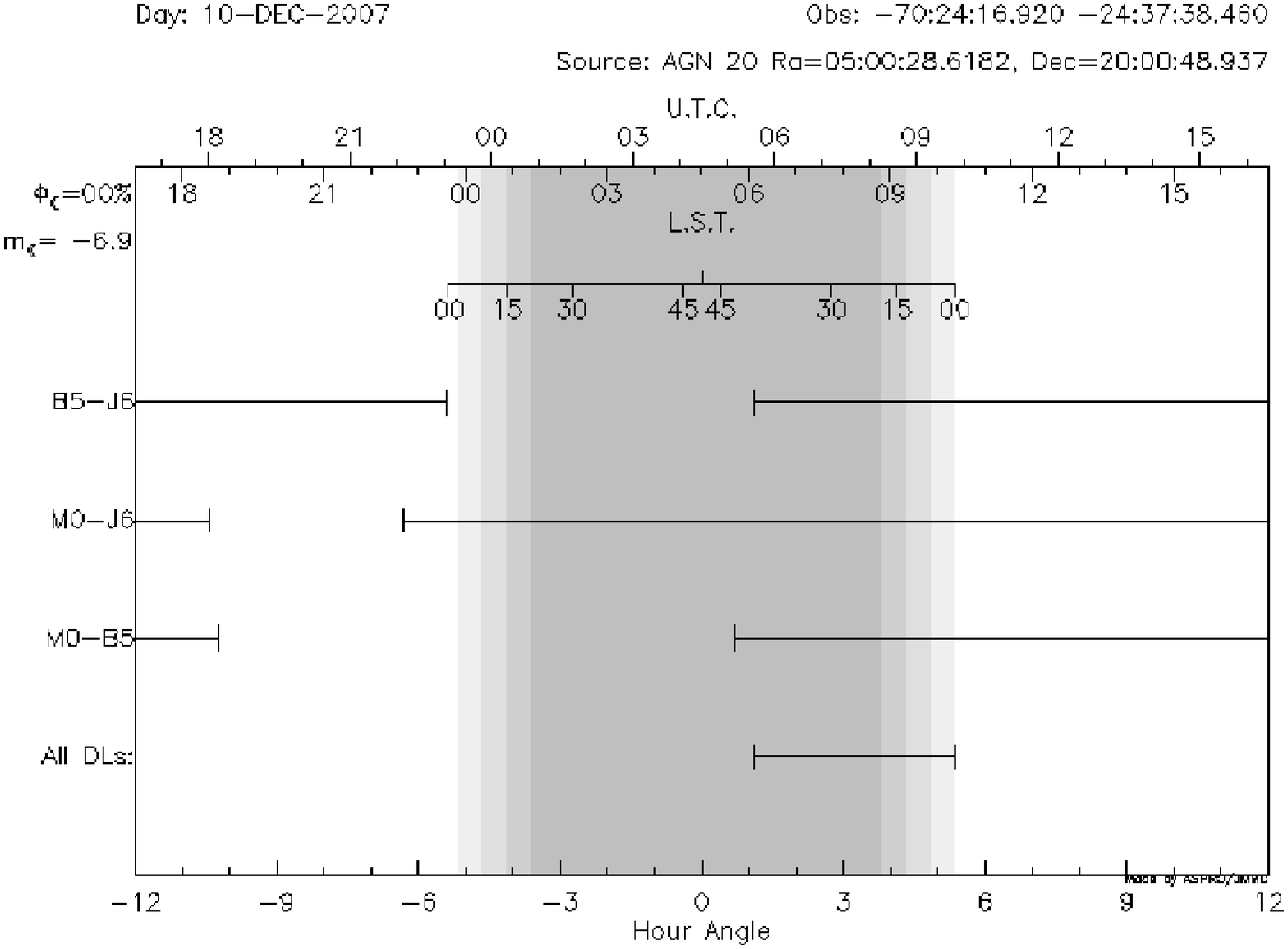}
  \end{tabular}
  \caption[Fu Ori disk model.]{
    \footnotesize{
      \emph{\bf Left:} Observability of sources for a big triangle of
      baselines (B5-J6-M0). The constraints due to the delay lines
      (red lines) are now larger than for 2 telescopes.
      \emph{\bf Right:} Detail of the delay line constraints: the
      B5-J6 and M0-B5 are the most constraining baselines.
    }
  }
  \label{fig:observability3Tels}
\end{figure}

\paragraph*{Observability:}
First, one needs to check the observability when taking a big triangle
(B5-J6-M0, Fig.~\ref{fig:observability3Tels}, left panel). As one can
see, the observability constraints are much more stringent than
before, using a lower number of telescopes. The \emph{Observability limits
  due to delay lines} panel (right side) now gives meaningful
information; i.e., splitting the constraint by delay line. Here, one can
see that the most constraining baselines are the B5-J6 and M0-B5, and
the M0-J6 baseline does not too much constrain the observability.

\begin{figure}[htbp]
  \centering
  \begin{tabular}{ccc}
    \multicolumn{3}{c}{\includegraphics[width=0.3\textwidth]{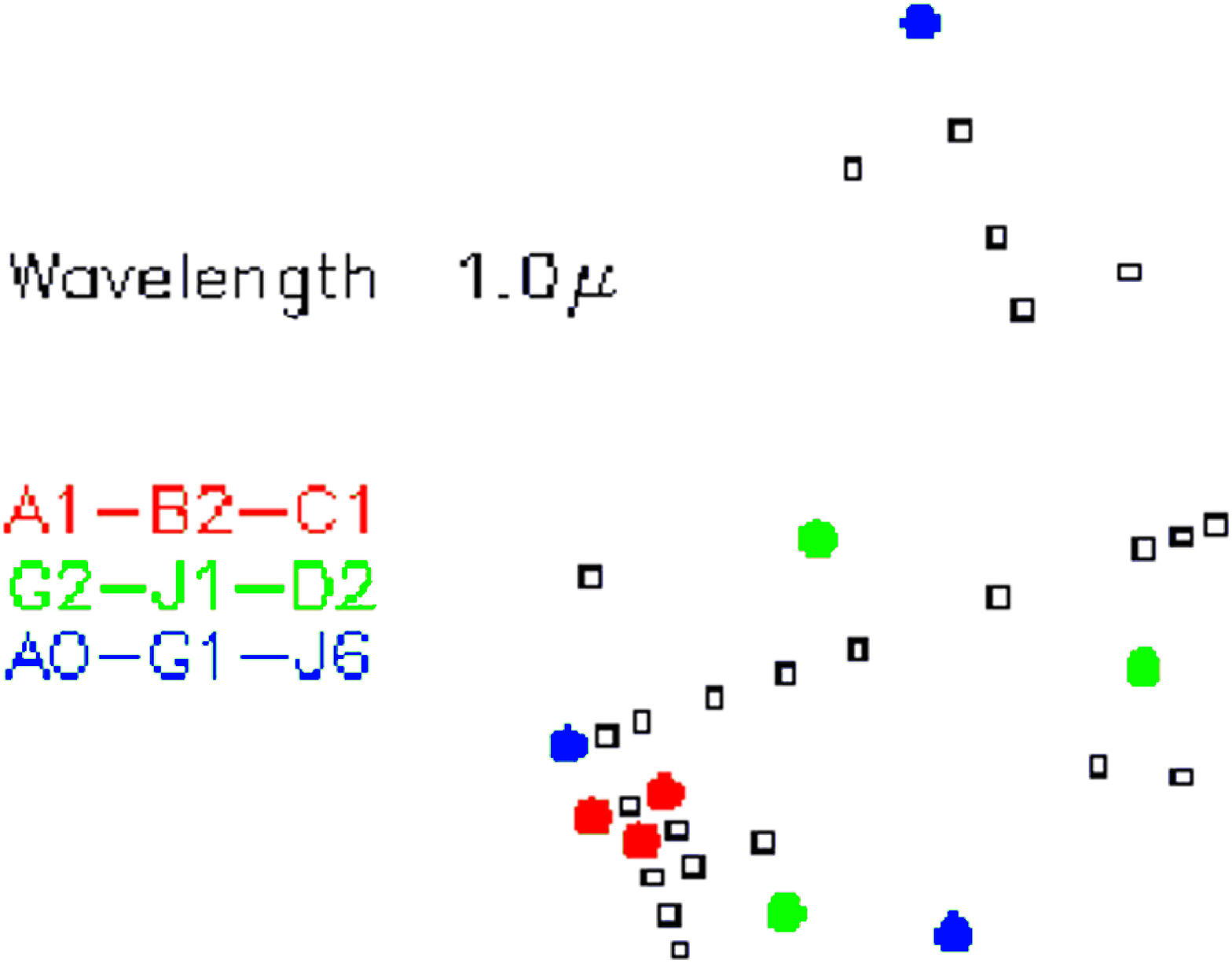}}\\
    \includegraphics[width=0.3\textwidth]{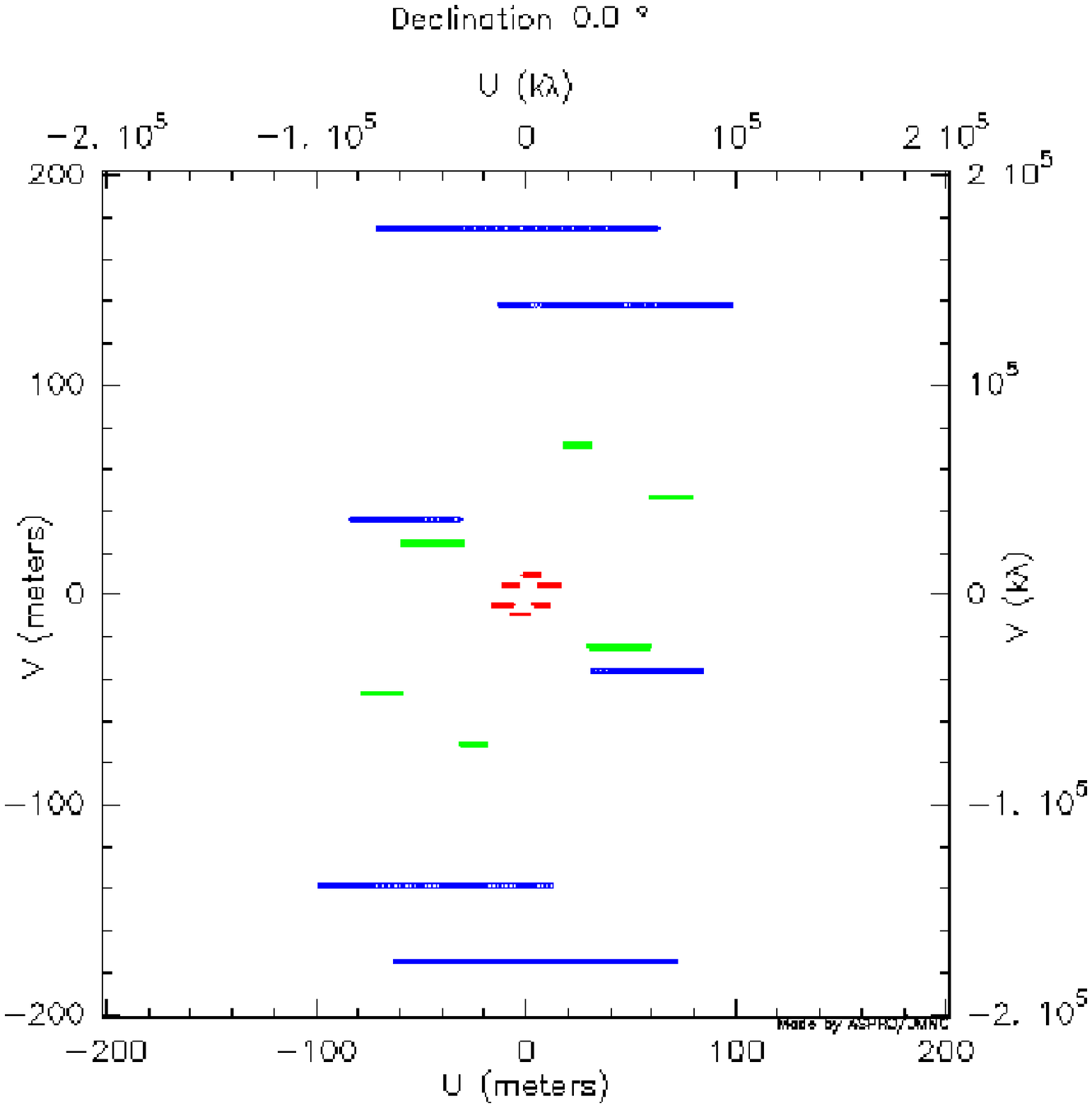}&
    \includegraphics[width=0.3\textwidth]{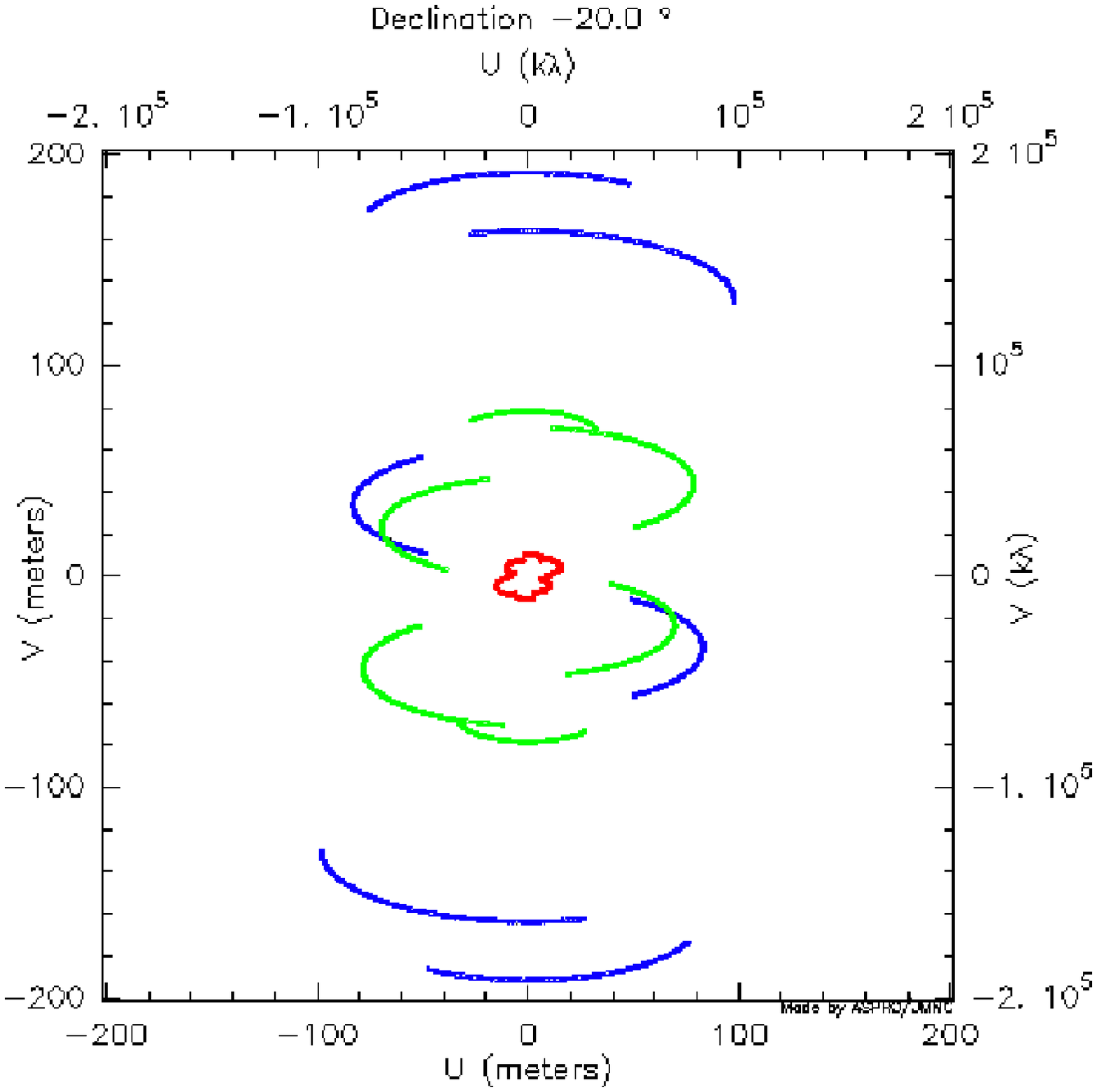}&
    \includegraphics[width=0.3\textwidth]{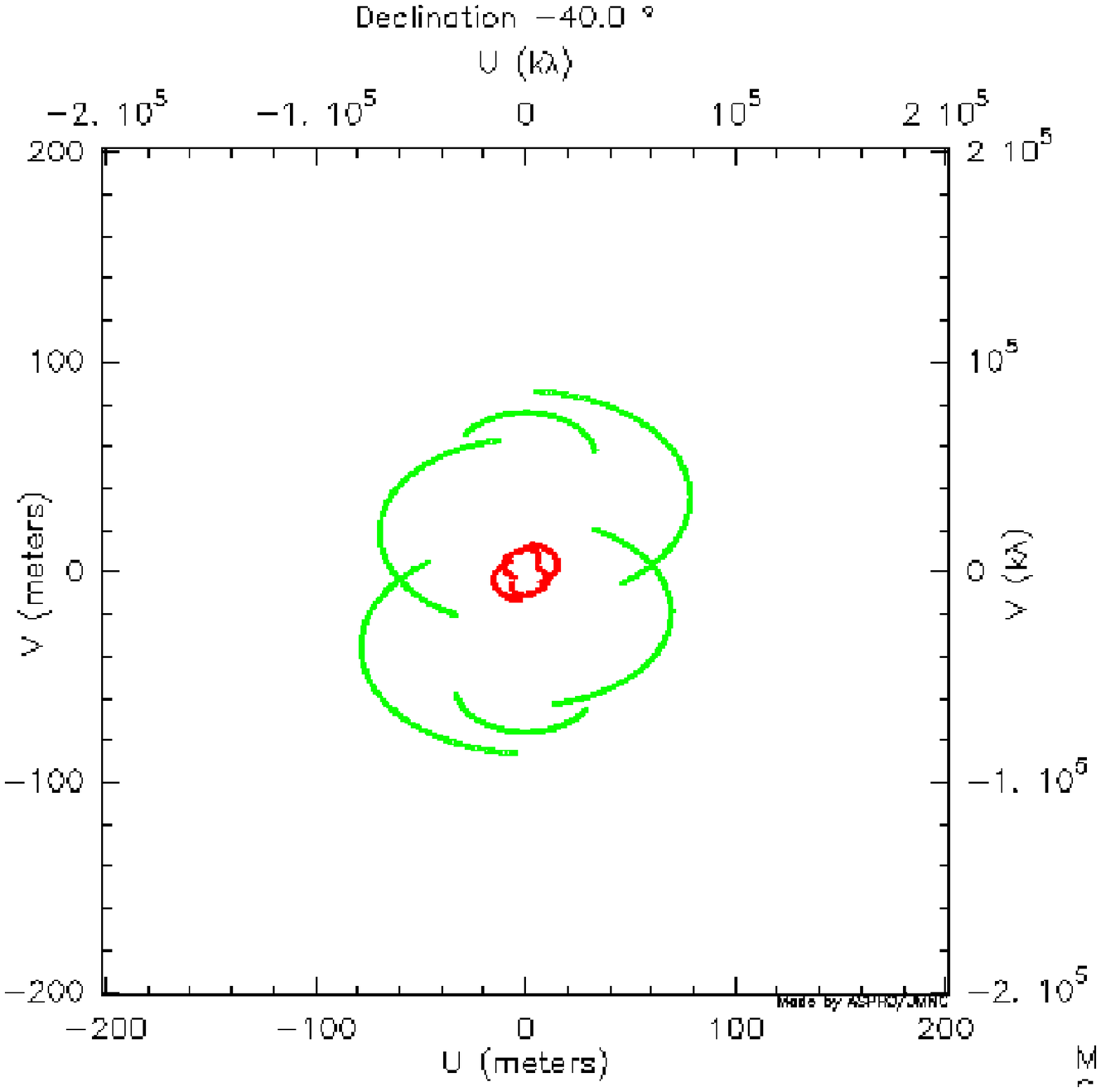}\\
    \includegraphics[width=0.3\textwidth]{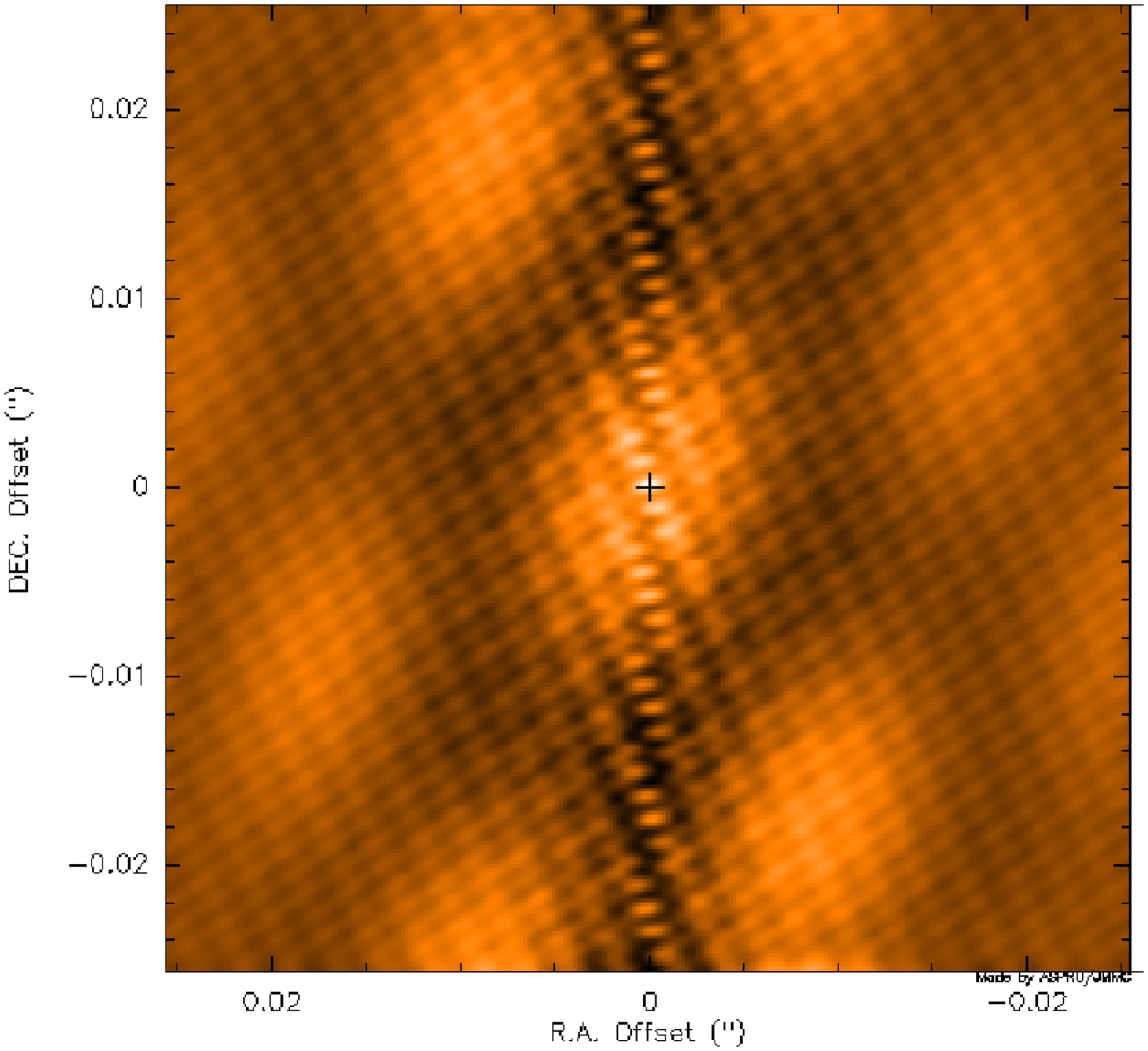}&
    \includegraphics[width=0.3\textwidth]{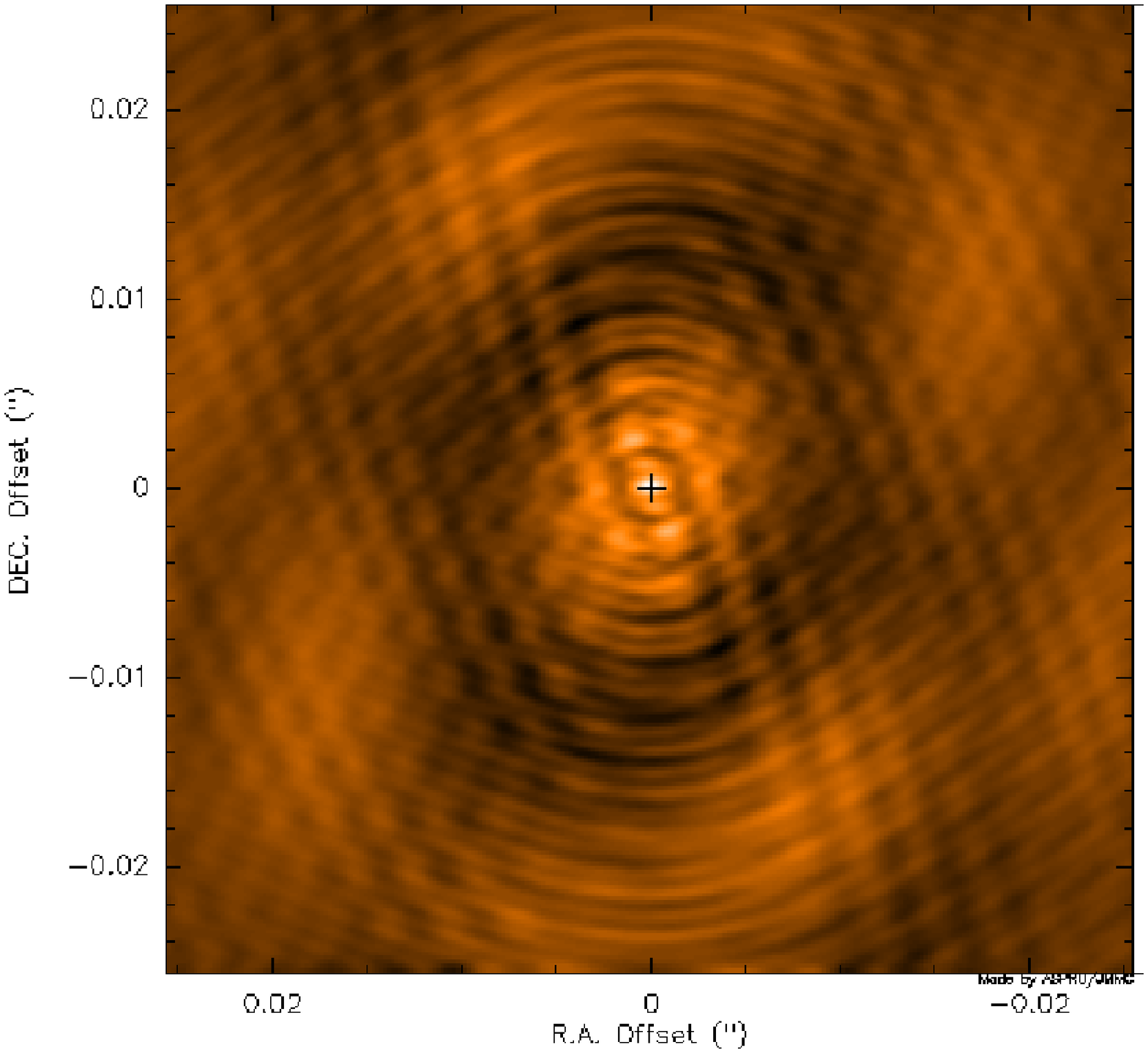}&
    \includegraphics[width=0.3\textwidth]{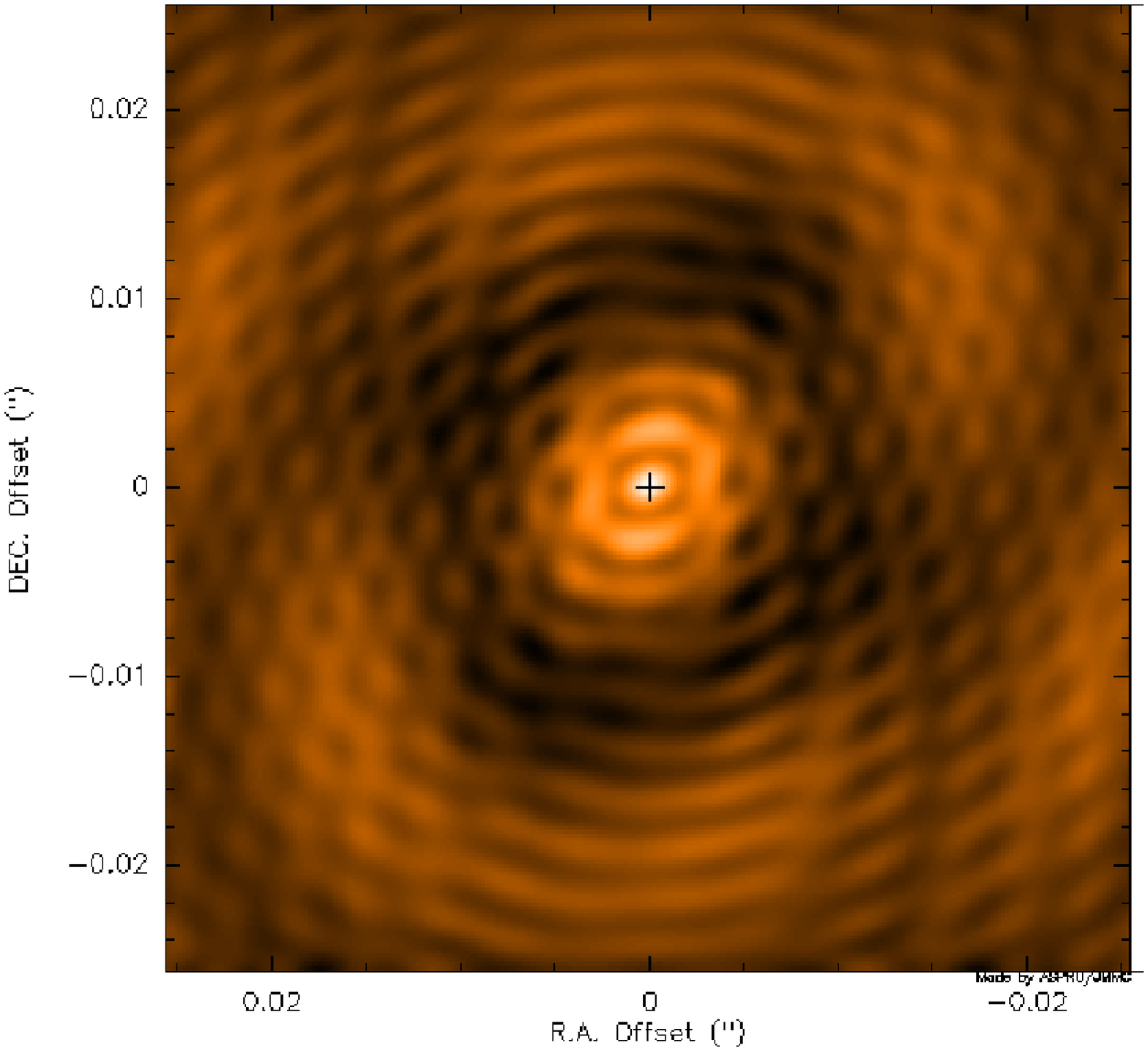}
  \end{tabular}
  \caption[Fu Ori disk model.]{
    \footnotesize{
      \emph{\bf Top:} Sketch of the stations used in this example.
      \emph{\bf Middle-Left:} $UV$ tracks for a 3-telescope observation on
      a 0\deg\ declination object and several telescopes
      configurations: A1-B2-C1, A0-G1-J6 and G2-J1-D2, corresponding
      roughly to 3 observing nights using AMBER and with complete
      freedom in the stations selection.
      \emph{\bf Bottom-Left:} The corresponding ``dirty beam'', or what
      would look like the PSF if one would observe a point-source with
      a telescope having an aperture similar to the previous $UV$ track.
      \emph{\bf Middle:} Same as 1$^{st}$ plot, but for a
      -20\deg\ declination target.
      \emph{\bf Bottom-Middle:} The corresponding dirty beam.
      \emph{\bf Middle-Right:} Same as before but for a -40\deg\
      declination target. Note that the biggest triangle does not
      appear in this plot, as the constraints due to the delay lines
      prevents from observing with these telescopes.
      \emph{\bf Bottom-Right:} The corresponding dirty beam.
    }
  }
  \label{fig:uvTracks3Tels}
\end{figure}

\paragraph*{$UV$ tracks for different configurations:} 
In Fig.~\ref{fig:uvTracks3Tels}, the $UV$ tracks for the
different triangles and different object declinations (0\deg, -20\deg\ 
and -40\deg\ from top to bottom) are shown. One can see that many aspects will
affect the $UV$ coverage:
\begin{itemize}
\item declination - which affects the quantity of different position
  angles.
\item delay lines constraints - which prevent one from observing with
  large baselines (bottom graph).
\item available baselines - which limits the available stations due to
  the interferometer possibilities.
\end{itemize}

\paragraph*{Beam shape:}  One can see that the many aspects raised
before will affect the $UV$ coverage and, therefore, the quality of the
PSF (plotted using the \emph{show dirty beam} option):
\begin{itemize}
\item  declination affects the quality of different position angles
  and gives irregular secondary lobes to the PSF.
\item The delay line constraints prevent one from observing with large
  baselines (bottom graph) and widen the PSF (loss of angular
  resolution).
\item The available baselines give an elongated PSF in the E-W
  direction for the VLTI, since the longest baseline is in the N-S
  direction. 
\end{itemize}

\subsection*{\underline{Exercise 6:} Radius measurement of a star (uniform disk)}

\begin{figure}[htbp]
  \centering
  \includegraphics[width=0.7\textwidth]{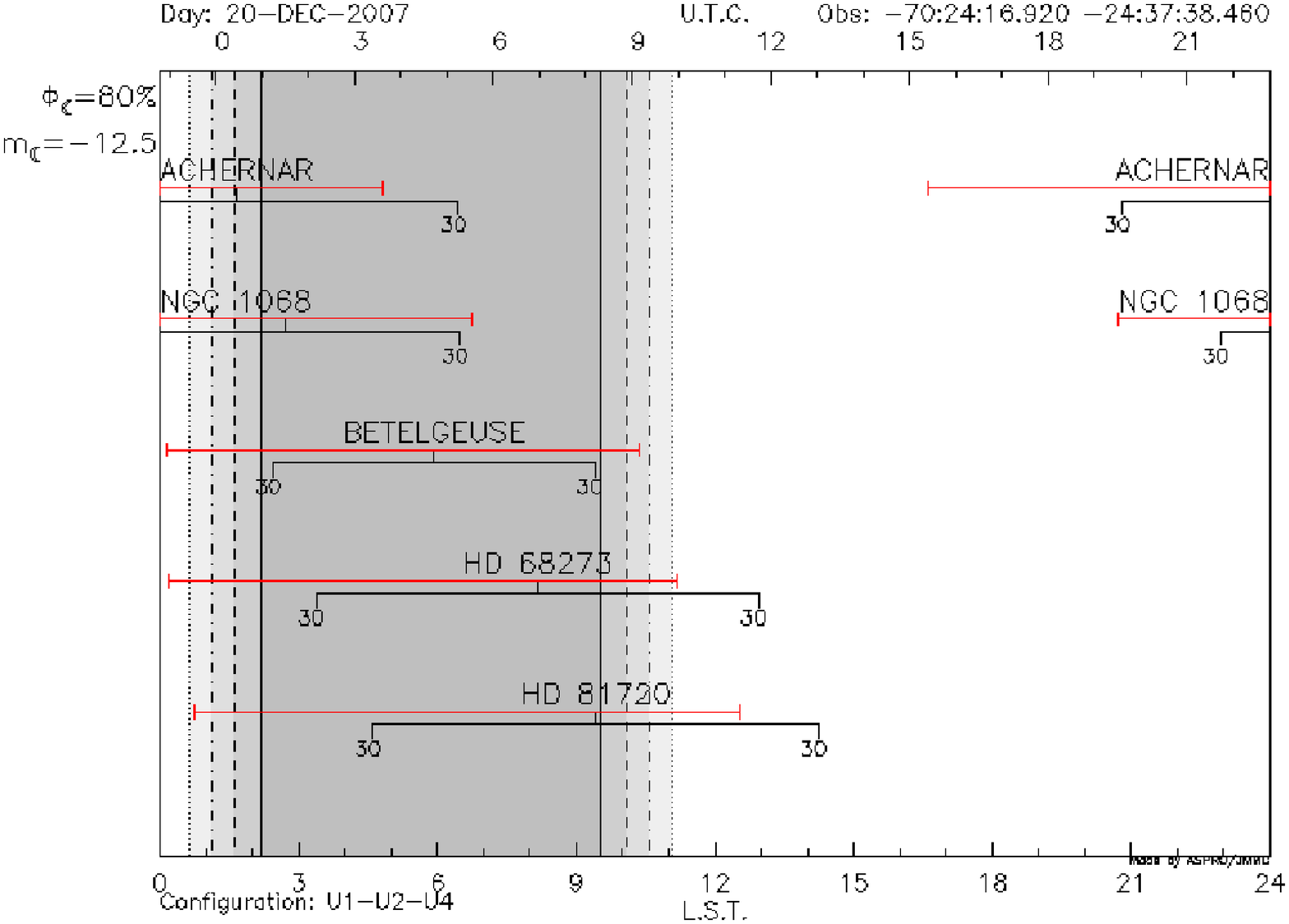}
  \caption[Fu Ori disk model.]{
    \footnotesize{
      Observability of the 1$^{st}$ catalog.
    }
  }
  \label{fig:obs1stCatalog}
\end{figure}

\paragraph*{Optimizing the observability of a series of sources:}
The first step is to set up an observation date in order to be able
to observe all targets together in one night. The optimal date of
observation would be the 15\th\ of December, but the moon is full on the
20\th. However, one can choose this date
(Fig.~\ref{fig:obs1stCatalog}) since interferometry is insensitive to
the moon phase. I also chose the UT1-UT3-UT4 triplet since there is a
faint target in my sample (NGC 1068) and checked that it does not put
too many additional constraints on the observability of the source due
to delay lines.

\paragraph*{Radius measurement:}
First of all, you need to look at the standard accuracy of the
instrument you will use to be able to know what you can expect from
your observations. The ESO \emph{Call for
  Proposal}\footnote{\url{http://www.eso.org/sci/observing/proposals/}}
gives you information about AMBER and MIDI (the two offered
instruments) accuracy you can expect. For AMBER in P81, the accuracy
is 3\% for an uncalibrated visibility point; i.e., about 5\% for a
calibrated one. If you look at the different targets here (I
have taken a diameter of 3 mas for NGC~1068, see
Fig.~\ref{fig:diamDetermination}), you can see that with
this accuracy, both the star Betelgeuse and HD~68273 ($\gamma^2$ Vel)
are unreachable for diameter measurement. The star HD~81720 will be
measurable but with a poor accuracy on the diameter. Therefore the
targets you will really be able to observe using these baselines are
Achernar, NGC~1068, and partially HD~81720. For the two other stars
(HD~68273 and Betelgeuse), you will need other configurations (a short
baseline triplet for Betelgeuse and the longest available baselines
for HD~68273) to reach your goal of measuring a diameter.

\paragraph*{More details about the object:} 
You should also notice that limb-darkening measurements, which need at
least one point in the second lobe of visibility, are not achievable,
except for Betelgeuse. Finally, one baseline setup is not sufficient
to measure all the star diameters: you will need at least a 3-baseline
setup to reach your proposal goal (one short baseline setup for Betelgeuse and
one very long baselines needed for HD~68273 and HD~81720 to
reach a better accuracy).

\paragraph*{Knowing the limitations:}
By browsing the ESO call for
proposals\footnote{\url{http://www.eso.org/sci/observing/proposals/}},
you can find that the current AMBER accuracy is 0.03 on the
raw visibilities. This means 0.05 on calibrated
measurements. Therefore, as seen in Fig.~\ref{fig:diamDetermination},
only Achernar and NGC1068 will allow one to ``easily'' measure a non-zero
or non-1 visibility. HD68273 and Betelgeuse will make the measurement
very difficult, as the expected visibility is very close to 0 or
1. Then, HD81720, whose visibility at maximum baseline is 0.85,
will be marginally resolved and only an upper limit to the diameter
will be measurable.

\begin{figure}[htbp]
  \centering
  \begin{tabular}{cc}
    \includegraphics[width=0.47\textwidth]{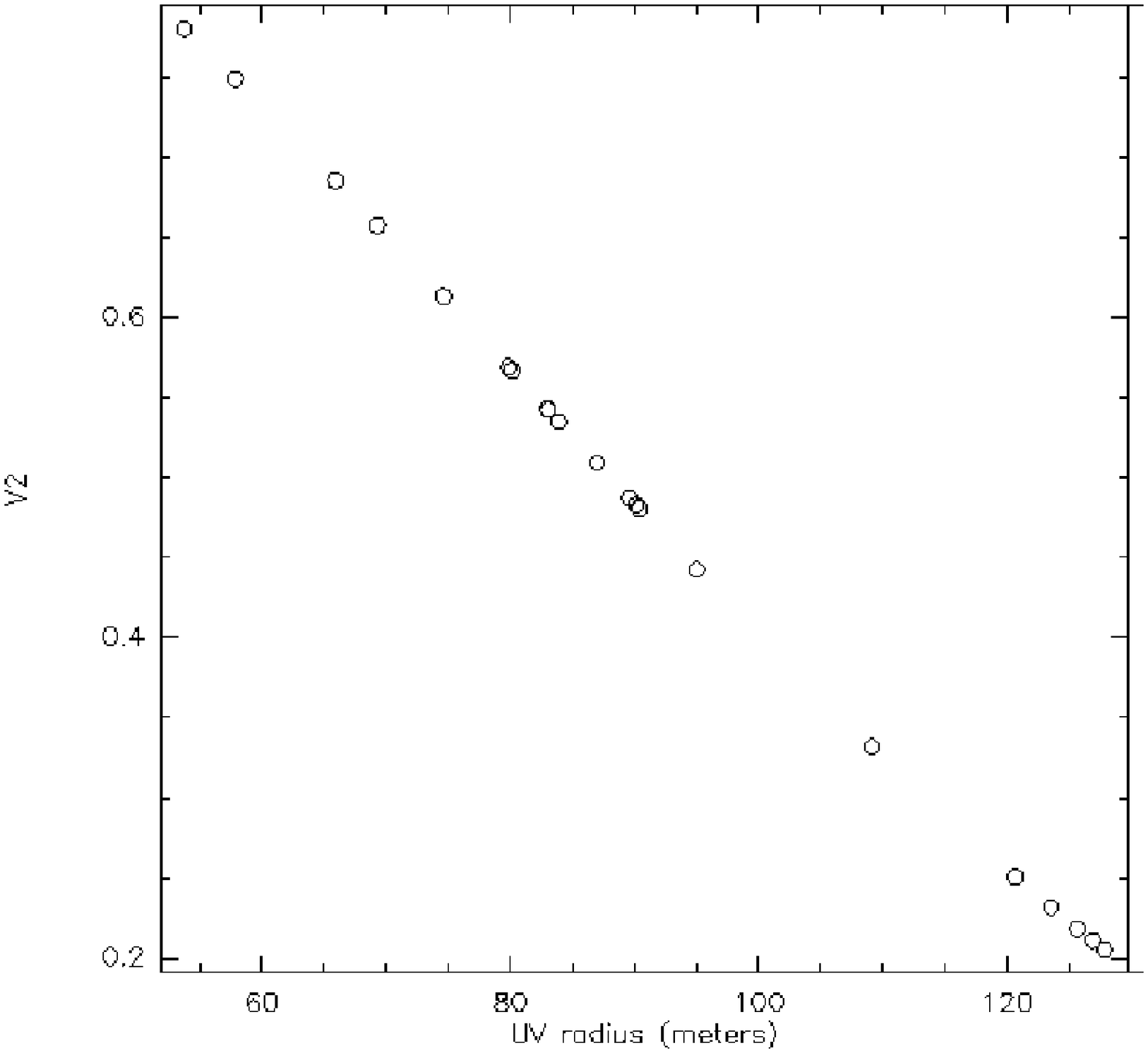}&
    \includegraphics[width=0.47\textwidth]{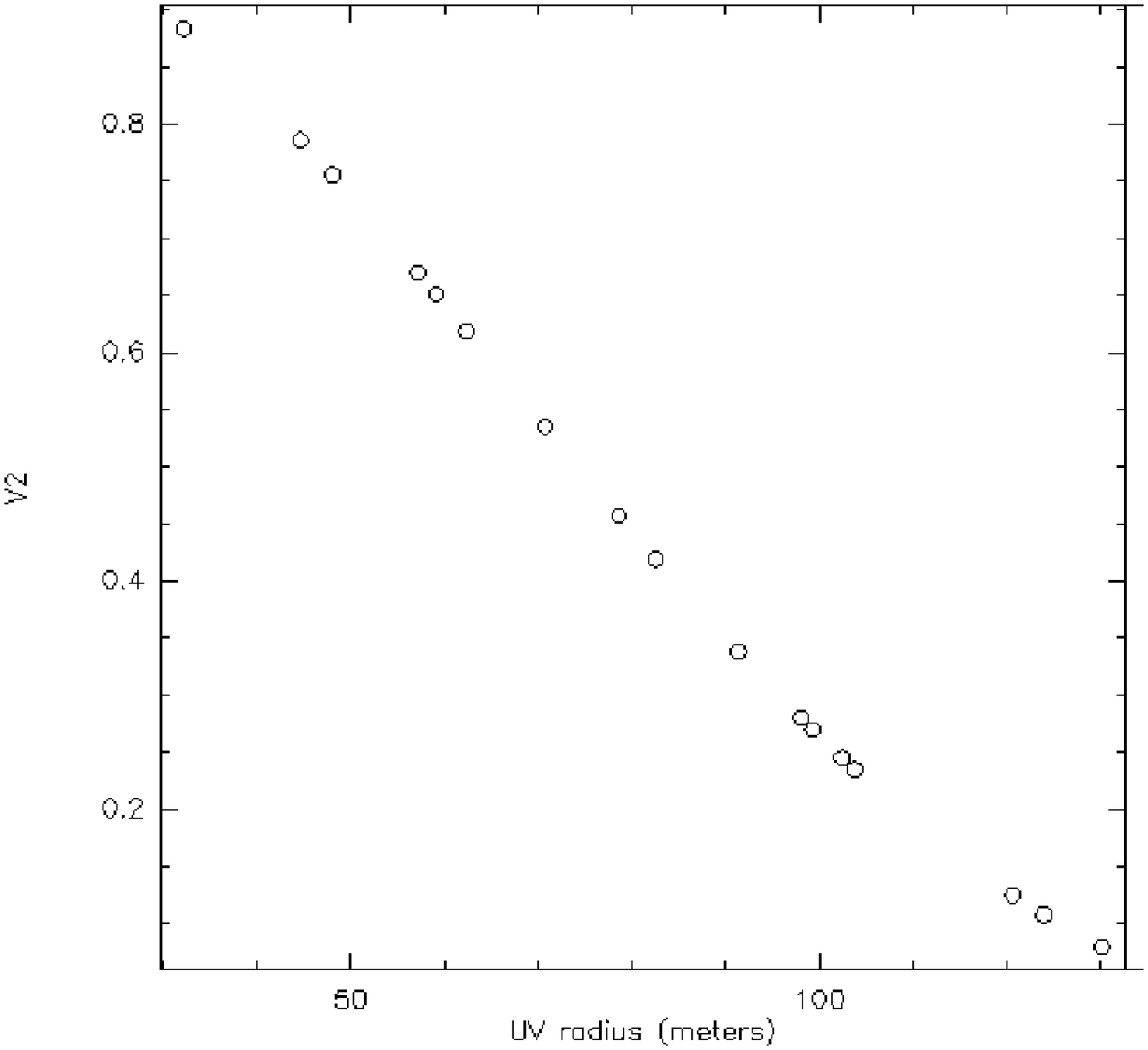}\\
    \includegraphics[width=0.47\textwidth]{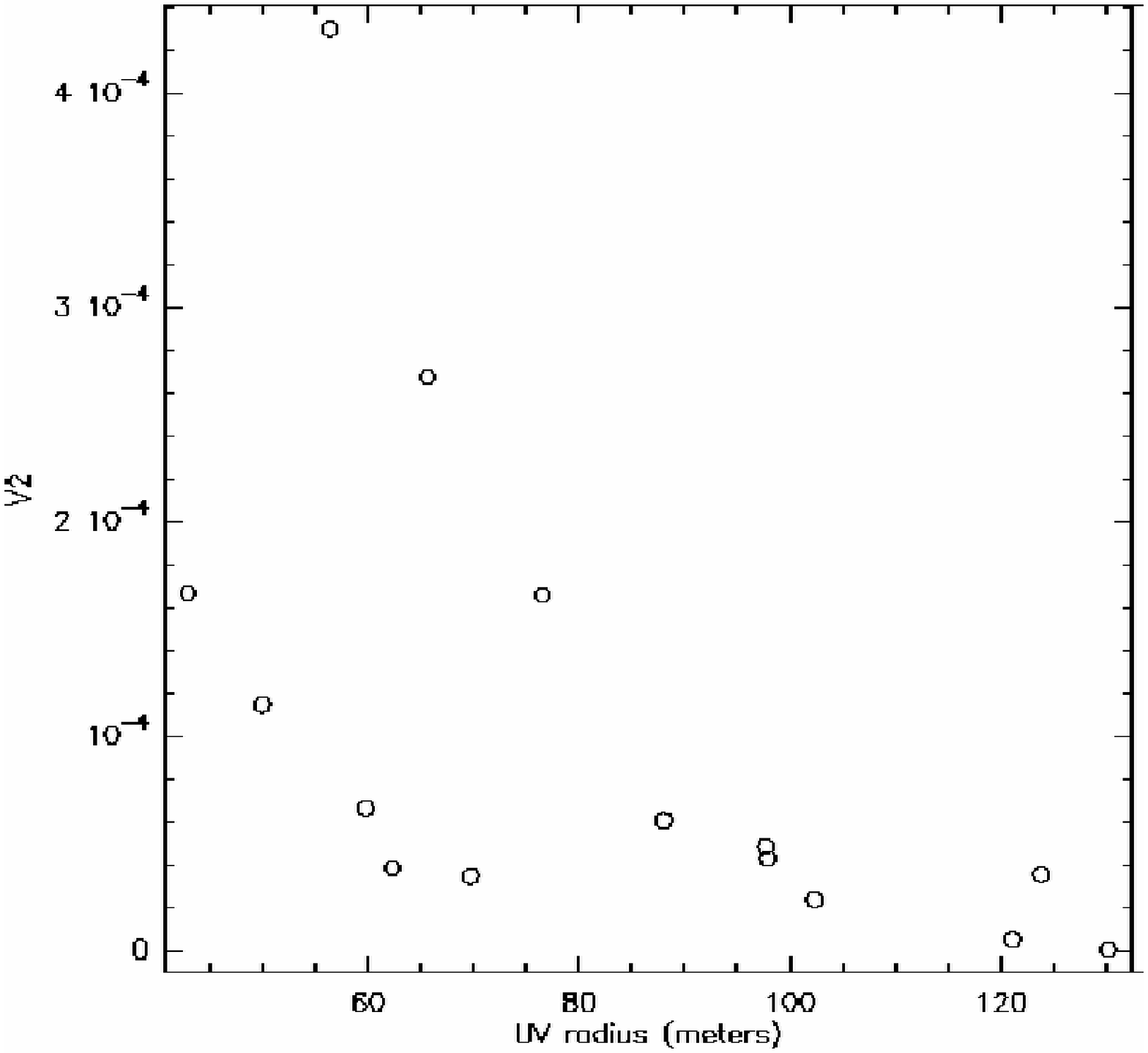}&
    \includegraphics[width=0.47\textwidth]{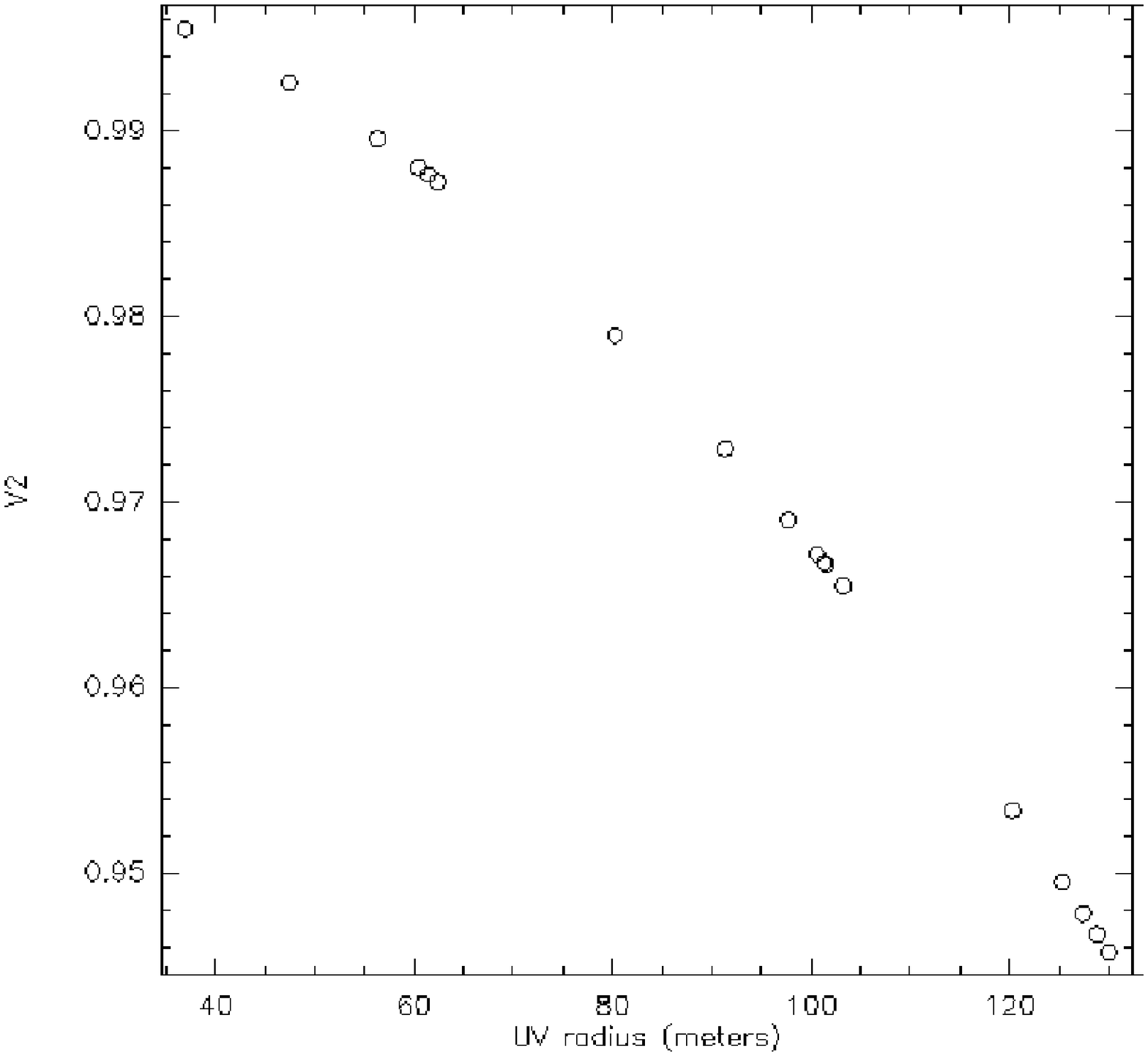}\\
    \includegraphics[width=0.47\textwidth]{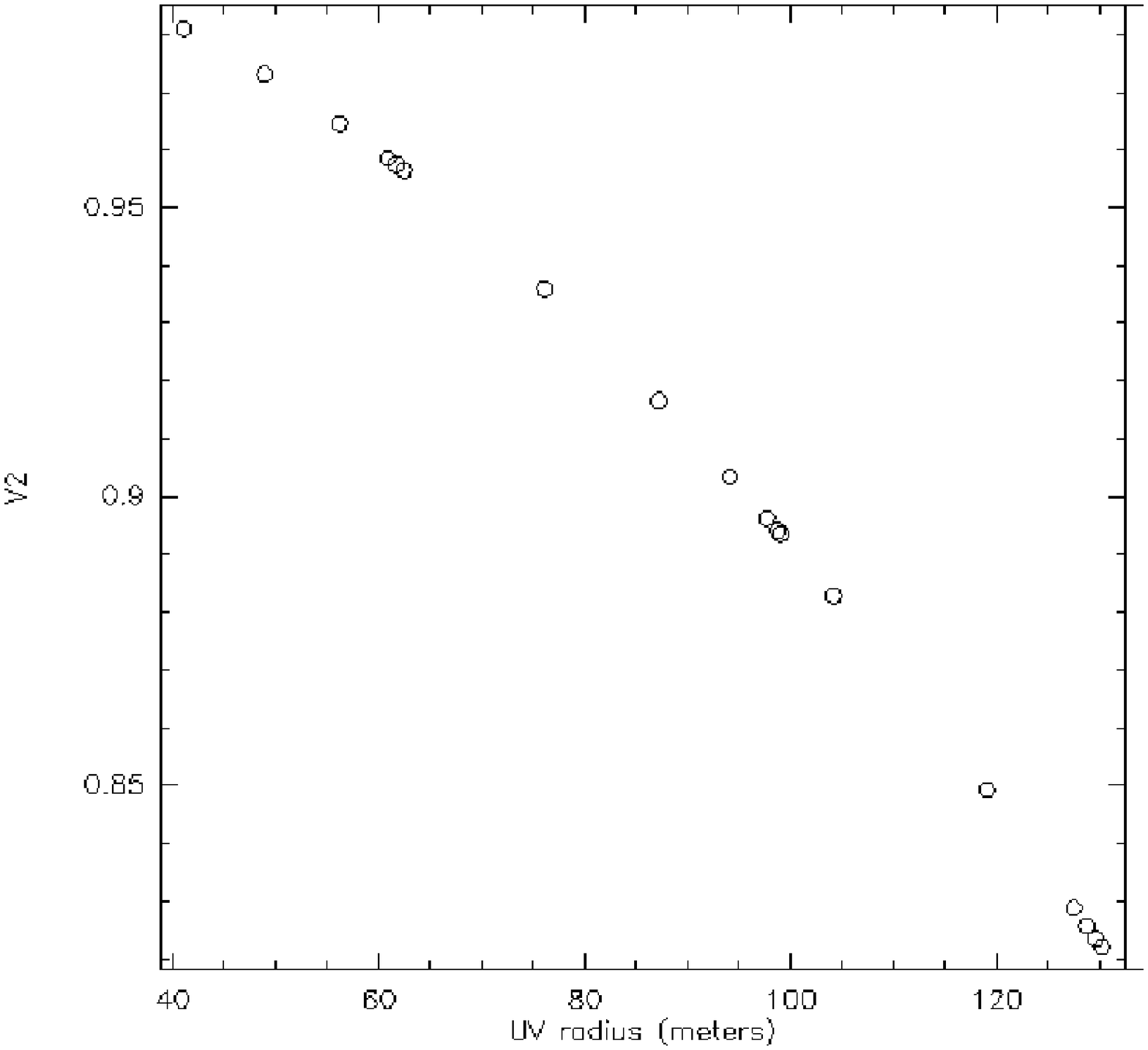}&
  \end{tabular}
  \caption[Fu Ori disk model.]
  {
    \footnotesize{
      \emph{\bf Top-Left:} Achernar visibilities for the largest
      baselines, given a radius of 2.53\,mas.
      \emph{\bf Top-Right:} Same for NGC 1068 and a diameter 3\,mas.
      \emph{\bf Middle-Left:} Same for Betelgeuse, diameter of
      44\,mas. 
      \emph{\bf Middle-Right:} Same for HD68273, diameter of 0.5\,mas.
      \emph{\bf Bottom-Left:} same for HD81720, diameter of 0.93\,mas.
    }
  }
  \label{fig:diamDetermination}
\end{figure}

\subsection*{\underline{Exercise 7:} Binary parameter determination}

If the reader has reached this point carried out all of the exercises
without any problems, he is now an expert in long-baseline stellar
interferometry and can do this exercise without any difficulty ;-)




\appendix
\section*{APPENDIX}
\section{Practical considerations}

\subsection{\texttt{ASPRO} rules of thumb}

\texttt{ASPRO} is an optical long-baseline stellar interferometry tool
intended to help the observations preparation. One can find it on the
website \url{http://www.jmmc.fr}. Note that all exercises can be done
at home with an internet connection, given that \texttt{ASPRO} can be
launched via a java web interface.

\paragraph*{Launching \texttt{ASPRO} on the web:} The \texttt{ASPRO}
launch is set up in five simple steps:
\begin{enumerate}
\item go to \url{http://www.jmmc.fr}
\item Select \emph{ASPRO applet}. A pop-up window should appear (if not,
  please allow pop-up windows from \url{http://www.jmmc.fr} in your
  web browser).
\item If you have a user account on the JMMC website, then just log in;
  otherwise, uncheck the \emph{start application using my account
    information} and proceed to the next step.
\item in the \emph{Start...} menu, select \emph{ASPRO}. 2 new windows
  should appear.
\item The last step is to select the \texttt{ASPRO} version you want
  to use in the menu \emph{Choose...}. Here we will use the \emph{Full
    ASPRO interface} version.
\end{enumerate}

\begin{figure}[htbp]
  \centering
  \includegraphics[width=0.9\textwidth]{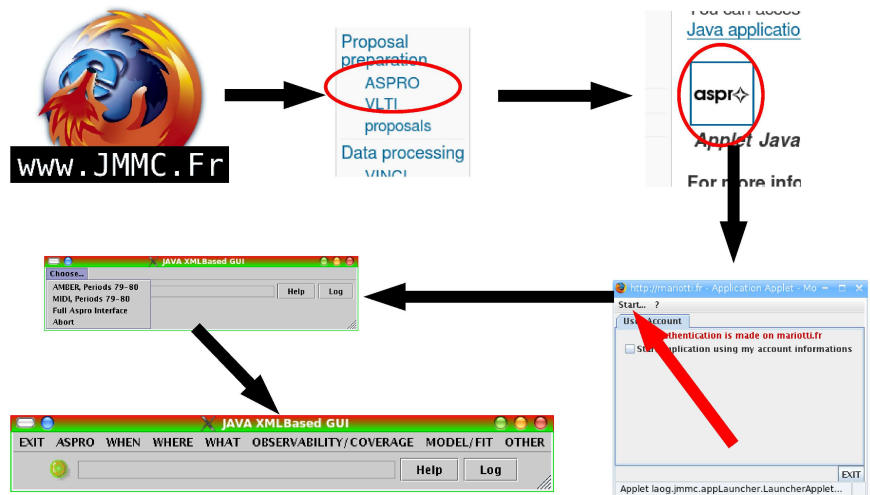}
  \caption[\texttt{ASPRO} on the web.]
  {
    \footnotesize{
      The different steps to launch the web version of \texttt{ASPRO}. One only
      needs an internet connexion and a java-enabled web browser.
    }
  }
  \label{fig:diamDetermination}
\end{figure}

\paragraph*{Launching \texttt{ASPRO} on a local computer:} If your
computer is one of the rare ones to have \texttt{ASPRO} locally
installed, then the launch is even simpler:
\begin{enumerate}
\item In a command line, type \texttt{aspro @oipt}
\item In the menu \emph{Choose...}, select the \texttt{ASPRO} version
  you want to use. Here, we will use the \emph{Full ASPRO interface}
  version.
\end{enumerate}

\begin{figure}[htbp]
  \centering
  \includegraphics[width=0.9\textwidth]{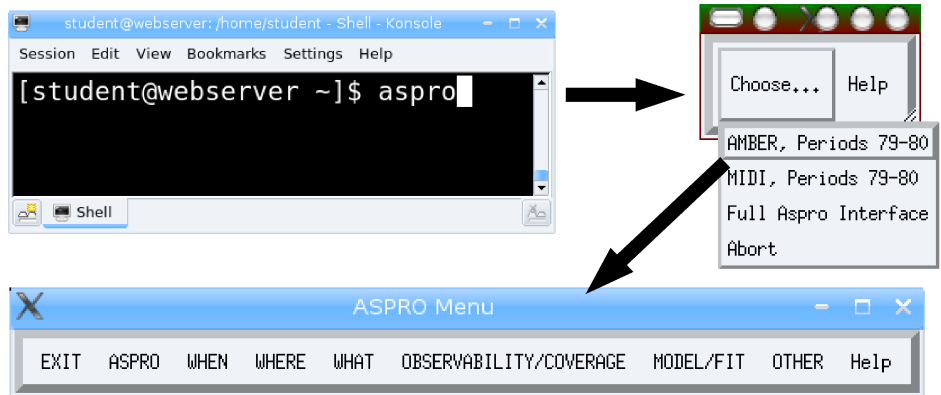}
  \caption[Fu Ori disk model.]
  {
    \footnotesize{
      The steps to launch \texttt{ASPRO} on a local computer: even
      simpler, but one needs a local version of the software
      installed!
    }
  }
  \label{fig:diamDetermination}
\end{figure}

\paragraph*{When \texttt{ASPRO} gets stuck:}

You will sometimes experience strange behavior such as non-responding
buttons or a different response to what you expect. If you are in doubt,
do not hesitate to quit \texttt{ASPRO} by clicking the \emph{EXIT}
button. If this button does not work either, you can still write
\texttt{exit} in the command line, which will kill the Gildas session
(and the \texttt{ASPRO} one at the same time).

Do not worry, restarting \texttt{ASPRO} and entering the different
parameters again does not take very long!


\begin{thebibliography}{}
\bibitem[Millour(2008)]{2008_Millour}
  Millour, F., All you ever wanted to know about optical long baseline
  stellar interferometry, but were too shy to ask your adviser, New
  Astronomy Review, this issue.
\end{thebibliography}
\end{document}